# Engineering high Pockels coefficients in thin-film strontium titanate for cryogenic quantum electro-optic applications

Anja Ulrich[1,2*], Kamal Brahim[1,3*], Andries Boelen[1,4*], Michiel Debaets[1,2], Conglin Sun[1,5], Yishu Huang[1,2], Sandeep Seema Saseendran[1], Marina Baryshnikova[1], Paola Favia[1], Thomas Nuytten[1], Stefanie Sergeant[1], Kasper Van Gasse[1,2], Bart Kuyken[1,2], Kristiaan De Greve[1,3], Clement Merckling[1,4], Christian Haffner[1†]

\* contributing equally, † corresponding author

1) Imec, B-3001 Leuven, Belgium
2) Department of Information Technology (INTEC), Photonics Research Group, Ghent University, B-9052 Ghent, Belgium
3) Department of Electrical Engineering (ESAT), KU Leuven, B-3001 Leuven, Belgium
4) Department of Materials Engineering (MTM), KU Leuven, B-3001 Leuven, Belgium
5) Department of Physics and Astronomy, KU Leuven, B-3001 Leuven, Belgium

## Summary

Materials which exhibit the Pockels effect are notable for their strong electro-optic interaction and rapid response times and are therefore used extensively in classical electro-optic components for data and telecommunication applications [1]. Yet many materials optimized for room-temperature operation [2-4] see their Pockels coefficients at cryogenic temperatures significantly reduced - a major hurdle for emerging quantum technologies which have even more rigorous demands than their classical counterpart [5,6]. A noted example is $BaTiO_3$, which features the strongest effective Pockels coefficient at room temperature, only to see it reduced to a third (i.e. $r_{eff} \approx 170$ pm/V) at a few Kelvin [3]. Here, we show that this behaviour is not inherent and can even be reversed: Strontium titanate ($SrTiO_3$), a material normally not featuring a Pockels coefficient, can be engineered to exhibit an $r_{eff}$ of 345 pm/V at cryogenic temperatures – a record value in any thin-film electro-optic material. By adjusting the stoichiometry, we can increase the Curie temperature and realise a ferroelectric phase that yields a high Pockels coefficient, yet with limited optical losses - on the order of decibels per centimetre. Our findings position $SrTiO_3$ as one of the most promising materials for cryogenic quantum photonics applications.

## Introduction – phase transitions, permittivity and nonlinearity

Quantum applications, including the transduction of microwave photons to optical frequencies [7,8] and the high-efficiency detection of photons [9] require cryogenic cooling of entire optical systems, including the electro-optic transducers. In the quantum domain, every photon counts – hence the electro-optic materials used need to provide quantum-imposed improved performance characteristics that are maintained at 4 K and below. The improved performance makes shorter electro-optic transducers possible, thus, minimizing photon loss. Said performance of the electro-optic material is best quantified by the Pockels coefficient ($r$) times the cube of the material's refractive index [10] ($n$) - ergo $n^3 \cdot r$. Together with the loss of the electro-optic material, it is the most important parameter to optimize for electro-optic systems. Figure 1 a) compares $n^3 \cdot r$ for various Pockels materials such as lithium niobate ($LiNbO_3$) [2,11,12], organic electro-optic chromophores (OEO) [4,13] and barium titanate ($BaTiO_3$) [3] at room temperature and cryogenic temperature. One inherent complication in the comparison arises from the fact that Pockels coefficients are tensorial, not scalar. To compare various materials, the indicated Pockels coefficient used here is an effective scalar value ($r_{eff}$) where the tensor elements are weighted by the strength of optical and radiofrequency (RF) fields.

Miller's empirical rule [14]: $r \propto (\varepsilon_r - 1) \cdot \delta$ illustrates how the strength of $r$ is intricately linked to a material's ability to break inversion symmetry at the atomic level ($\delta \neq 0$). Thereby it significantly enhances its linear polarizability ($\varepsilon_r$-1), linked to the material's permittivity $\varepsilon_r$, by a factor $\delta$ (hyperpolarizability) [14]. From Miller's rule, two practical strategies can be employed to engineer materials with high Pockels coefficients: either work with materials exhibiting a high hyperpolarizability and a non-negligible linear polarization, or engineer materials to obtain large linear polarizations and non-zero hyperpolarization. As part of the first strategy chromophores can be used which feature a large hyperpolarizability via the breaking of electron cloud symmetry; this is achieved by attaching strong acceptor and donor groups to both ends of a π-bridge, while the linear polarization remains in the single digits [14]. Organic chromophores achieve Pockels coefficients of up to 140 pm/V at 4 K [4], only ~ 10% lower than room temperature values. To achieve such large hyperpolarizability, the π-bridge needs to be a few nanometers long. This results in rather narrow bandgaps ~ 1 eV that cause absorption losses, typically ≥1 dB/cm [13,14], and imposes its own challenges.

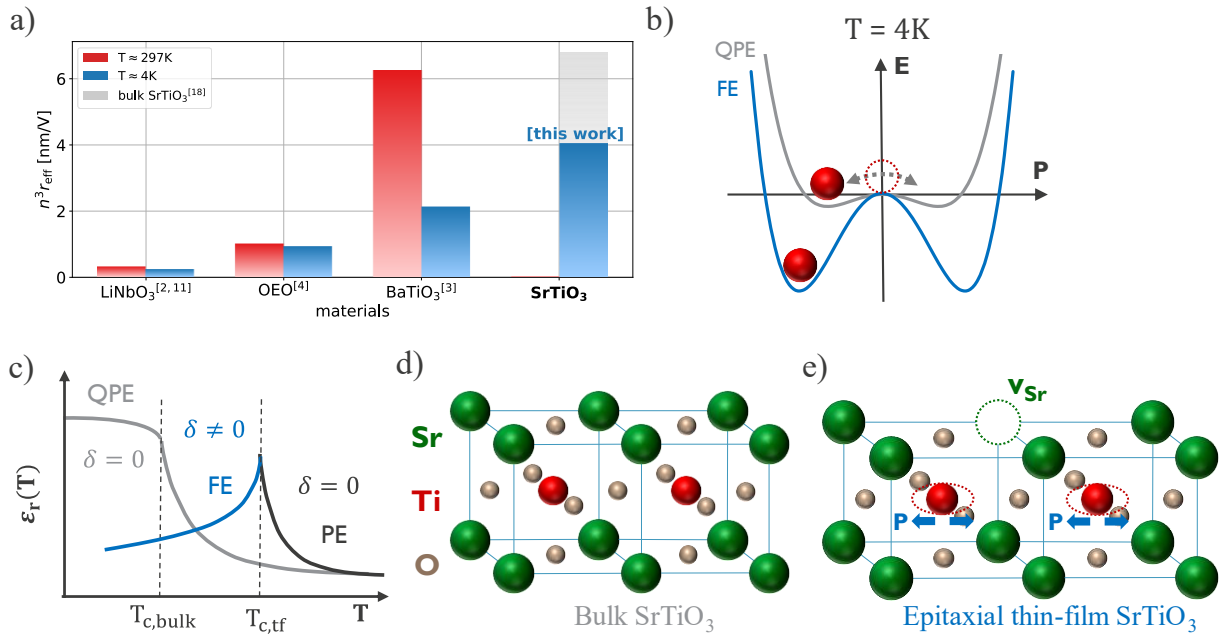

Figure 1 a) Reported effective electro-optic strength of known thin-film Pockels materials on silicon dioxide such as lithium niobate (LiNbO$_3$), organic chromophores (OEO), barium titanate (BaTiO$_3$) and this papers strontium titanate (SrTiO$_3$) at room temperature and around 4K [Supplementary 5]. SrTiO$_3$ features the largest coefficient, highlighting its potential for cryogenic applications. b) Energy vs. polarization potential for the Ti atom showing the difference in potential for quantum paraelectric (QPE) and ferroelectric (FE). In the former case, quantum fluctuations prevent the localization of the Ti atom, and the formation of a proper ferroelectric phase transition. c) Permittivity vs temperature schematics showing the effect of mode softening around the Curie temperature (T$_C$) for the thin-film (tf) case (FE) and bulk SrTiO$_3$ case (QPE). The hyperpolarizability is given for the different cases indicating symmetric ($\delta = 0$) or broken symmetry ($\delta \neq 0$) phases. d) Centrosymmetric crystal structure of bulk SrTiO$_3$ at 4 K. e) Inversion symmetry broken crystal structure of thin-film SrTiO$_3$ with displaced Ti atoms corresponding to the two energy minima shown in the ferroelectric potential in b).

As part of the second strategy, ferroelectric complex oxides break symmetry by displacing atoms on angstrom scales – resulting in a non-zero but significantly lower hyperpolarizability. Instead, their electro-optic response is driven by a large permittivity in the RF spectrum, originating from material resonances such as space-charge regions, dipolar relaxation, and phononic lattice resonances. However, only the latter extend all the way into the GHz frequency range – providing the bandwidth needed for quantum applications. In general, complex oxides feature both better material stability and lower optical losses compared to organic chromophores.

Therefore, the question becomes how can such permittivity engineering be realized? From symmetry configurations, ferroelectric phases are expected to be required for non-zero hyperpolarizability. When cooling down around the paraelectric to the ferroelectric phase transition, the phononic modes tend to soften resulting in a larger permittivity. Hence, by working slightly below but 'close' to the Curie temperature ($T_C$) in the ferroelectric phase, a high permittivity as shown in Figure 1 c) and thus a strong electro-optic effect could be expected. This engineering must also balance the tradeoff with losses naturally occurring at the transition (Kramer-Kronig [10]). Of the known complex oxides, $LiNbO_3$ is a prime Pockels material example in optical communications due to its very low (sub-decibel per meter) absorption loss and proven reliability [12]. However, due to the relative stiffness of the phononic mode of $LiNbO_3$ ($T_C \approx 1200$ K), its resonant enhancement is rather limited, and the resulting linear permittivity of 30 remains rather moderate, resulting in an electro-optic coefficient of 24 pm/V at cryogenic temperatures – quite a bit lower than those of e.g. the chromophores [2]. In contrast, $BaTiO_3$ features a $T_C$ of 400 K offering a high linear permittivity of a few thousand at room temperature – boosting $r_{\text{eff}}$ to $520 \pm 20$ pm/V [3]. Unfortunately, this does not persist at cryogenic temperatures as the phononic modes stiffen and $BaTiO_3$ also undergoes several crystal phase transitions when cooling down [15]. This results in a lower linear permittivity and thus reduced Pockels coefficient of $r_{\text{eff}} \approx 170 \pm 20$ pm/V [3].

For cryogenic applications, materials with lower $T_C$ close to the operation temperature might seem suitable; however, this at first appears impractical. Strontium titanate ($SrTiO_3$) with a $T_C$ of $\approx 37$ K, exemplifies inherent issues with low-$T_C$ materials due to quantum fluctuations that hinder the paraelectric to ferroelectric phase transition [16]. The net result being a quantum paraelectric material that, until now, was not expected to be a Pockels material. Figure 1 b) illustrates this behavior based on the energy potential of the central Ti atom. In the quantum paraelectric phase, the ground state of the Ti atom is not localized within the two potential minima. This results in a high polarizability as the Ti atom can be strongly displaced by weak fields. However, the lack of a broken symmetry ($\delta = 0$) forces the Pockels coefficient to be zero.

In this study, we challenge this conventional understanding by engineering thin-film $SrTiO_3$ to break its symmetry. $SrTiO_3$ was epitaxially grown on silicon with subsequent wafer bonding onto a silicon dioxide substrate. The slight Ti-rich flavor of the film in combination with strain increases the $T_C$ to 100 K, *high* enough to negate the effect of quantum fluctuations and thereby enable a proper, symmetry-breaking phase, see Figure 1 e). Yet $T_C$ is still *low* enough to benefit from the high permittivity due to lattice resonances. Under this condition our thin-film $SrTiO_3$ features an electro-optic strength of $r_{\text{eff}} \approx 345 \pm 30$ pm/V at 4 K. In addition, dedicated epitaxial growth and subsequent annealing optimization of the thin films result in low optical losses -- on the order of decibels per centimeter, which is in contrast to previous reports and conventional assumption that associates thin-film $SrTiO_3$ layers with high optical losses [17]. Recently in isotopically doped quantum paraelectric $SrTiO_3$ a large quadratic electro-optic Kerr effect was observed under DC bias [18]; with an equivalent Pockels coefficient of 1150 pm/V. Our study on wafer-scale thin-film $SrTiO_3$ in combination with the large quadratic electro-optic Kerr effect observed in the bulk crystal underscores the potential of $SrTiO_3$ to surpass traditional Pockels materials in electro-optic performance.

## Engineered symmetry breaking and ferroelectricity

High-quality epitaxial SrTiO$_3$ films were grown slightly Ti-rich by molecular beam epitaxy (MBE) on 200 mm (001)-oriented Si wafers [19]. Rutherford backscattering spectroscopy (RBS) determined a stoichiometry of Sr/Ti = 0.96 ± 0.03. We expect this slightly Ti-rich stoichiometry to provide a good trade-off between symmetry-breaking [20] while still preserving a high crystalline structure, -- the latter imposing growth conditions around Sr/Ti = 1. Post-growth, the wafer was wafer-bonded onto a low refractive index SiO$_2$ box to optically isolate the SrTiO$_3$ film from the high refractive index silicon substrate [Supplementary 1]. Additionally, high-temperature oxygen annealing treatments were performed to reduce the oxygen vacancy concentration and further increase the film's crystallinity [21]. The thin-film crystallinity was quantified via X-ray diffraction (XRD) and benchmarked against commercial references by comparing the full width at half maxima (FWHM) data of the ω-scan from the SrTiO$_3$ (002) diffraction peak. Figure 2 a) shows the FWHM wafer map featuring an average FWHM of 0.11 ± 0.02. In comparison, XRD analysis of 16 commercial bulk SrTiO$_3$ (001) substrates (1 x 1 cm$^2$) showed a FWHM of 0.022 ± 0.007°. Further to this transmission electron microscopy (TEM) analysis confirmed the epitaxial nature of the SrTiO$_3$ layer, revealing a sharp SiO$_2$/SrTiO$_3$ interface, see Figure 2 b). Threading dislocations, which are common in heteroepitaxial thin films, were observed to extend vertically through the full SrTiO$_3$ layer; from the substrate interface to the top surface. These can act as space charge regions, disrupt lattice uniformity, and increase conductivity.

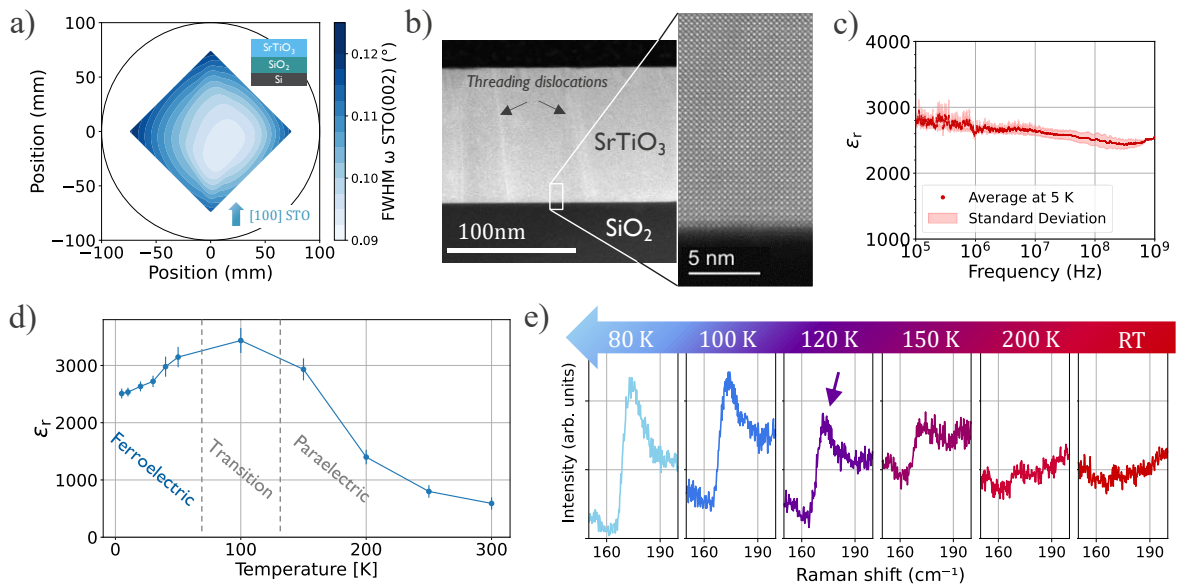

Figure 2 a) FWHM of the SrTiO$_3$ (002) ω-scan mapped over a full 200 mm wafer, showing good uniformity of the crystalline quality. b) Cross-sectional TEM of the epitaxial 105 nm SrTiO$_3$ thin film. The inset highlights the film crystallinity. c) Relative permittivity ($\varepsilon_r$) vs frequency response from 100 kHz to 1 GHz at 5 K, showing a large but slowly decreasing permittivity which drives the strong Pockels coefficient over a broad frequency range up to 1 GHz. d) Temperature dependence from 5 K to 300 K of SrTiO$_3$'s permittivity at 100 MHz. Around 100 K a paraelectric to (non-centrosymmetric) ferroelectric phase transition occurs resulting in the symmetry breaking required for a Pockels coefficient to exist. e) Temperature-dependent Raman spectroscopy of the SrTiO$_3$ thin film. Upon cooling down from room temperature (RT), the Raman signal at ∼ 173 cm$^{-1}$ appears. This is the phononic lattice resonance which corresponds to vibrational displacements of the Ti and O atoms, indicating a symmetry breaking.

Impedance spectroscopy was employed to directly measure the large cryogenic RF permittivity. These measurements were further extended to encompass a broad temperature range from 5 K to 300 K, and for frequencies spanning from 100 kHz to 1 GHz, using specially designed interdigitated capacitors. Upon increasing the temperature from 5 K to room temperature, the permittivity peaks at 100 K before dropping to ~ 600 at room temperature, see Figure 2 d). Such behavior is normally associated with a paraelectric (at high temperature) to ferroelectric (low temperature) phase transition around $T_C$. The phase transition breaks the crystal symmetry and gives rise to a non-zero Pockels coefficient. While pure ferroelectric materials typically feature a sharp phase transition at $T_C$, our $SrTiO_3$ film demonstrates a broader continuous transition from the paraelectric to ferroelectric phase during cooldown. Furthermore, we observe a slight lowering of permittivity with frequency from 2700 at 100 kHz to 2500 at 1 GHz Figure 2 c). Such broadening of the phase transition and reduction in permittivity over frequency is commonly associated with relaxor ferroelectric behavior - linked to polar nanoregions [22,23]. Polarizing and depolarizing these polar nanoregions adds a contribution to the relative permittivity at low frequencies. Our measurement indicates that this low-frequency effect is limited to ≈ 10% of the total permittivity. As the Pockels coefficient scales with the permittivity, our data [Figure 2 c)] suggests that the high Pockels coefficient of $SrTiO_3$ can be sustained up to GHz operations.

To verify the $SrTiO_3$ ferroelectric phase transition, temperature-dependent Raman spectroscopy was performed to measure the phononic lattice modes, see Figure 2 e). For $SrTiO_3$ in the (quantum) paraelectric phase, first-order Raman scattering (involving only one phonon) is symmetry forbidden as the unit cell is centrosymmetric. Upon cooling, a peak at ~ 173 cm$^{-1}$ appears, which corresponds to lattice modes, in particular to the longitudinal component of the so-called Slater mode and transversal component of the Last mode, both oscillating at 5.2 THz. The Slater mode describes mutually opposite vibrational displacements of the Ti and O atoms along the c-axis (axis of Ti displacement), whereas the Last mode corresponds to the coupled vibrations of the $TiO_6$ octahedra and the Sr ions [22,23]. The (gradual) appearance of this peak below 200 K indicates a (gradual) breaking of the crystal centrosymmetry, further supporting our observations of relaxor ferroelectric behavior [24]. Room-temperature lattice parameters of a = 3.905 ± 0.005 Å (in-plane) and c = 3.901 ± 0.005 Å (out-of-plane) suggest a cubic like $SrTiO_3$ unit cell that mimics bulk properties ($a_{bulk}$ = 3.905 Å) [Supplementary 2]. However, due to a factor three difference in thermal expansion coefficients (CTE) between thin film and substrate, cooling down the sample can induce a strain of ~ 0.18%. This can contribute to the aforementioned $SrTiO_3$ ferroelectric phase transition upon cooling [25]. Conversely, the resulting strain may still be partially relaxed due to dislocations which are able to release strain within the film, hence further investigation is needed to estimate the effective $SrTiO_3$ strain at cryogenic temperatures.

## Electro-optic interaction strength and bandwidth

The symmetry breaking phase of $BaTiO_3$ or $SrTiO_3$ is accomplished by the displacement of the Ti atom within a unit cell. Naturally, this displacement can occur in any direction, and its randomness results in a net zero macroscopic electro-optic effect. Applying a poling field aligns the atomic displacement and results in a macroscopic electro-optic response. To investigate the poling behavior of our thin films and quantify the electro-optic response strength, a Mach-Zehnder interferometer (MZI) was designed. We used polymethyl methacrylate (PMMA) loaded waveguides in combination with chromium/platinum electrodes to measure the electro-optic properties of $SrTiO_3$ at 4 K. PMMA and the $SiO_2$-substrate can be considered electro-optically inert, ensuring that the observed effects are primarily due to the 43% light confinement within the SrTiO$_3$ layer, see Figure 3 a). The measured waveguide propagation losses are dominated by coupling leakage to substrate modes; quantifying these, we estimate that the material losses of our SrTiO$_3$ films are below 5.5 dB/cm [Supplementary 6]. Applying an electric field ($E_{AC}$) induces a phase change of the light propagating in the bottom waveguide ($\Delta \psi \propto v_{pol} \cdot r_{eff} \cdot E_{AC}$). Here, $v_{pol}$ describes the degree of Ti atoms being displaced in the same direction and can reach values of ±1 for perfect alignment. The designed MZI, depicted in Figure 3 b), was used to translate the electric field induced phase changes into a modulation of the optical power by means of destructive and constructive interference between light propagating in the modulated and unmodulated arm. Figure 3 c) shows the output power over time (bottom) for a 20 Hz sinusoidal poling field with 1.6 V/μm amplitude (top). The strong oscillation of power indicates the presence of a large electro-optic effect, in detail a voltage change of 0.084 V/μm results in a π phase shift – this corresponds to a $V_\pi L \approx 1.04 \pm 0.08$ Vcm [Supplementary 4.2].

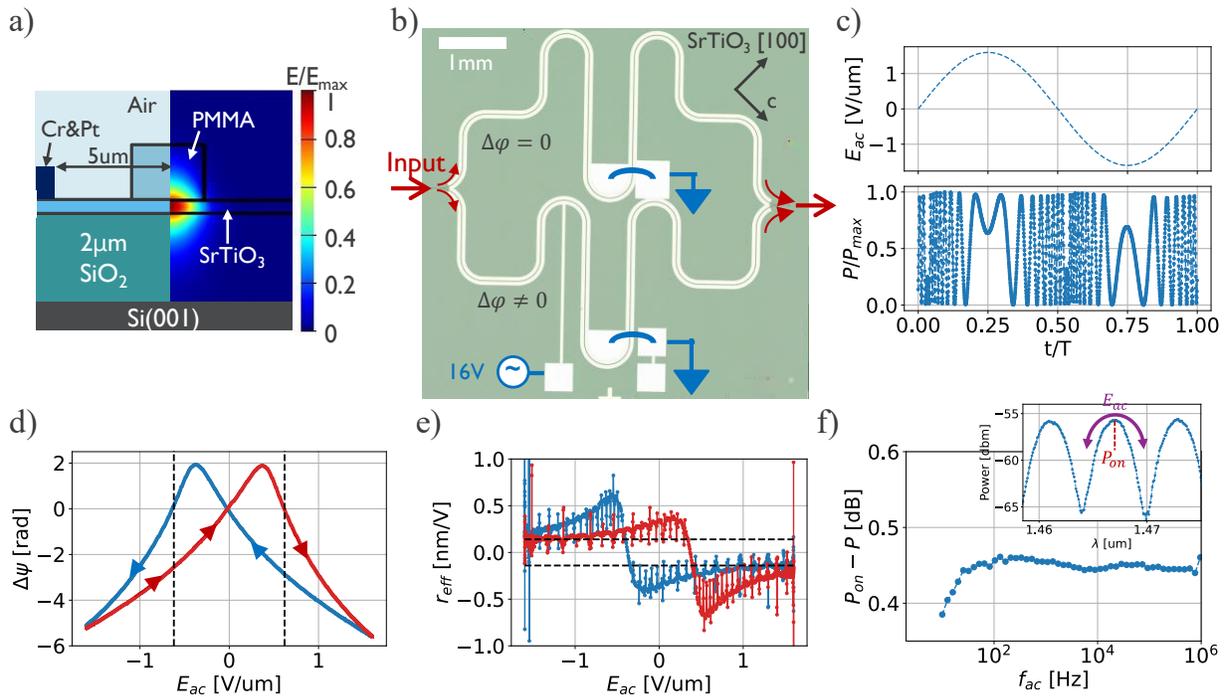

Figure 3 a) (left) Schematic cross section of the device, including (right) the simulated optical mode profile. b) Top view of the meandered MZI with the ferroelectric crystal structure indicated by the black arrows. The bottom arm is modulated by an electrical signal and destructively/constructively interferes with the upper arm, allows to track the phase change by measuring the transmitted power. c) (top) One period of the applied 20 Hz, 1.6 V/μm drive signal and (bottom) the optical output power response of the MZI. d) The induced phase change in the lower MZI arm vs applied electric field, indicative of so-called poling behavior, where the blue and red color corresponds to the down and up sweep of the voltage, respectively. Vertical dashed lines mark the coercive field $E_c = 0.6 \pm 0.05$ V/μm due to hysteretic behavior, which confirms ferroelectricity. e) Pockels coefficient $r_{eff}$ extracted from d). The poled value of 140 ± 30 pm/V at 1.6 V/μm is indicated by dashed horizontal lines. The reported 345 ± 30 pm/V in the main text corresponds to the (remanent) zero field value in the up direction (red). f) Electro-optic bandwidth measurement (10 Hz to 1 MHz) as sketched in the inset. An AC signal causes a drop of the average optical power per ac oscillation cycle (purple arrow) depending on the electro-optic strength. For frequencies below 20 Hz the response drops – this likely originates from a potential DC screening effect.

The extracted phase versus applied field [Supplementary 4.2] is depicted in Figure 3 d) and shows a hysteresis when sweeping the voltage up (red) and down (blue). This is indeed indicative of a ferroelectric phase in the thin-film SrTiO$_3$ [26]. The vertical dashed lines in the poling curve of Figure 3 d) highlight the coercive field ($E_C$) of 0.6 ± 0.05 V/µm at which enough energy is provided to switch back previously poled domains to generate a net zero phase change. For fields significantly higher than the coercive field, the poling function becomes a constant and $r_{eff}$ in Figure 3 e) should flatten out. However, under strong fields a drop of $r_{eff}$ can be observed [Supplementary 4.2] which we attribute to a ~ 60% lowering of the permittivity under strong fields. This behavior is in line with what has been reported for ferroelectrics [23]. Therefore, the achievable electro-optic coefficient for future applications will depend on the operation conditions (i.e. bias, ac voltages, modulation speed) and is shown here to reach values of up to 345 ± 30 pm/V under small bias, dropping to 140 ± 30 pm/V with a 1.6 V/µm bias field. The latter value was confirmed by performing a 1 kHz small signal modulation [Supplementary 4.3]. Quantum paraelectric $SrTiO_3$ shows a similar trend of lower electro-optic strength when applying a bias field > 0.1 V/µm [18]. Therefore, devices based on SrTiO3 in the ferroelectric phase might be better suited for applications that require high bias voltages to keep optical losses at a minimum [5].

Due to bandwidth limitations in our cryogenic setups, we used an indirect small-signal method to infer the electro-optic frequency response. This method uses the effect that an applied AC signal has on the average optical power, resulting in a drop with respect to MZI's on-point ($P_{on}$), see inset of Figure 3 e). This drop is proportional to the electro-optic strength of the Pockels material. Figure 3 e) shows that the drop in optical power is unaffected by an increase in ac modulation frequency indicating a constant modulation up to 1 MHz [Supplementary 4.4]. The electro-optic bandwidth measurement in combination with the up to 1 GHz permittivity measurements indicates the absence of sub-GHz resonances and promises high-frequency operation of $SrTiO_3$. The reduced response at low frequencies (< 50 Hz) prevented any DC bias experiments [Supplementary 4.5]. We speculate that this low-frequency response could be due to an oxygen vacancy related mechanism which effectively screens bias fields. Whether this is related to non-stochiometric $SrTiO_3$ in the vicinity of the electrodes or oxidization of the Cr- $SrTiO_3$ interface [27] remains to be investigated in follow-up studies.

## Outlook

In this work, we have demonstrated for the first time the presence of a record high Pockels effect in thin-film $SrTiO_3$ at 4 K, driven by its ferroelectric properties. Future research is needed to fully explore how well this material might be suited for various applications, and to determine where its fundamental limitations lie. In this sense, our studies provide a first glance into the world offered by an active electro-optic $SrTiO_3$, with clear routes to further optimization through material engineering, e.g. by leveraging stoichiometry, strain and/or isotope doping [18]. Dedicated engineering optimization will also be crucial to enable further reduction of optical losses for $SrTiO_3$ to reach its full potential. Like many other novel photonic materials, the capability to integrate $SrTiO_3$ with existing photonics platforms based on silicon nitride or silicon will be crucial to provide practical relevance. And while lithium niobate poses contamination risks in large-scale semiconductor and photonics fabrication, $SrTiO_3$ is known to present fewer contamination challenges-- derisking the challenging integration road ahead. While the observed high Pockels coefficients are promising, the full Pockels tensor needs to be disentangled, and the exact nature of the RF losses remains an open question that needs to be addressed before this material can be considered in a quantum transducer to bridge superconducting computing and optical fiber communications.

## References

1. Boes, A. *et al.* Lithium niobate photonics: Unlocking the electromagnetic spectrum. *Science* **379**, eabj4396 (2023).
2. Zhu, D. *et al.* Integrated photonics on thin-film lithium niobate. (2021) doi:10.48550/ARXIV.2102.11956.
3. Eltes, F. *et al.* An integrated optical modulator operating at cryogenic temperatures. *Nat. Mater.* **19**, 1164–1168 (2020).
4. Bisang, D. *et al.* Plasmonic Modulators in Cryogenic Environment Featuring Bandwidths in Excess of 100 GHz and Reduced Plasmonic Losses. *ACS Photonics* **11**, 2691–2699 (2024).
5. Aghaee Rad, H. *et al.* Scaling and networking a modular photonic quantum computer. *Nature* (2025) doi:10.1038/s41586-024-08406-9.
6. Moody, G. *et al.* 2022 Roadmap on integrated quantum photonics. *J. Phys. Photonics* **4**, 012501 (2022).
7. Witmer, J. D. *et al.* A silicon-organic hybrid platform for quantum microwave-to-optical transduction. *Quantum Sci. Technol.* **5**, 034004 (2020).
8. Mirhosseini, M., Sipahigil, A., Kalaee, M. & Painter, O. Superconducting qubit to optical photon transduction. *Nature* **588**, 599–603 (2020).
9. Elshaari, A. W., Pernice, W., Srinivasan, K., Benson, O. & Zwiller, V. Hybrid integrated quantum photonic circuits. *Nat. Photonics* **14**, 285–298 (2020).
10. Boyd, R. W. *Nonlinear Optics*. (Academic Press, Burlington, MA, 2008).
11. Herzog, C., Poberaj, G. & Günter, P. Electro-optic behavior of lithium niobate at cryogenic temperatures. *Opt. Commun.* **281**, 793–796 (2008).
12. Shams-Ansari, A. *et al.* Reduced material loss in thin-film lithium niobate waveguides. *APL Photonics* **7**, 081301 (2022).
13. HLD - Organic Non-linear Optical (NLO) Material with High Electro-optic Effects. (2021).
14. Dalton, L. R., Günter, P., Jazbinsek, M., Kwon, O.-P. & Sullivan, P. A. *Organic Electro-Optics and Photonics: Molecules, Polymers and Crystals*. (Cambridge University Press, 2015). doi:10.1017/CBO9781139043885.
15. Karvounis, A., Timpu, F., Vogler-Neuling, V. V., Savo, R. & Grange, R. Barium Titanate Nanostructures and Thin Films for Photonics. *Adv. Opt. Mater.* **8**, 2001249 (2020).
16. Müller, K. A. & Burkard, H. $SrTiO_3$ : An intrinsic quantum paraelectric below 4 K. *Phys. Rev. B* **19**, 3593–3602 (1979).
17. Eltes, F. *et al.* Low-Loss $BaTiO_3$ –Si Waveguides for Nonlinear Integrated Photonics. *ACS Photonics* **3**, 1698–1703 (2016).
18. Anderson, C. P. et al. Quantum critical electro-optic and piezo-electric nonlinearities. Preprint at https://doi.org/10.48550/arXiv.2502.15164 (2025).
19. Boelen, A. *et al.* Stoichiometry and Thickness of Epitaxial $SrTiO_3$ on Silicon (001): an Investigation of Physical, Optical and Electrical Properties. Preprint at https://doi.org/10.48550/arXiv.2412.07395 (2024).
20. Kang, K. T. *et al.* Ferroelectricity in $SrTiO3$ epitaxial thin films via Sr-vacancy-induced tetragonality. *Appl. Surf. Sci.* **499**, 143930 (2020).
21. Baryshnikova, M. *et al.* Impact of Cationic Stoichiometry on Physical, Optical and Electrical Properties of $SrTiO_3$ Thin Films Grown on (001)-Oriented Si Substrates. *Materials* **17**, 1714 (2024).

22. Dwij, V., De, B. K., Tyagi, S., Sharma, G. & Sathe, V. Fano resonance and relaxor behavior in Pr doped SrTiO3: A Raman spectroscopic investigation. *Phys. B Condens. Matter* **620**, 413265 (2021).
23. Wördenweber, R., Schubert, J., Ehlig, T. & Hollmann, E. Relaxor ferro- and paraelectricity in anisotropically strained $SrTiO_3$ films. *J. Appl. Phys.* **113**, 164103 (2013).
24. Linnik, E. D. *et al.* Raman Response of Quantum Critical Ferroelectric Pb-Doped SrTiO3. *Crystals* **11**, 1469 (2021).
25. Haeni, J. H. *et al.* Room-temperature ferroelectricity in strained $SrTiO_3$. *Nature* **430**, 758–761 (2004).
26. Ma, Z. *et al.* Modeling of hysteresis loop and its applications in ferroelectric materials. *Ceram. Int.* **44**, 4338–4343 (2018).
27. Weilenmann, C. *et al.* Single neuromorphic memristor closely emulates multiple synaptic mechanisms for energy efficient neural networks. *Nat. Commun.* **15**, 6898 (2024).
28. Hoshina, T., Sase, R., Nishiyama, J., Takeda, H. & Tsurumi, T. Effect of oxygen vacancies on intrinsic dielectric permittivity of strontium titanate ceramics. *J. Ceram. Soc. Jpn.* **126**, 263–268 (2018).
29. Luo, Y. *et al.* Asymmetric Mach–Zehnder interferometer-based optical sensor with characteristics of both wavelength and temperature independence. *J. Opt.* **52**, 1008–1021 (2023).
30. Haffner, C. *et al.* All-plasmonic Mach–Zehnder modulator enabling optical high-speed communication at the microscale. *Nat. Photonics* **9**, 525–528 (2015).

Methods

### Growth

Heteroepitaxial SrTiO$_3$ thin films were grown on Si (001)-oriented substrates using MBE. The p-type Si (001) wafers were cleaned for 90 s in a 2% HF solution to remove part of the organic residues from the surface prior to introducing them to the ultra-high vacuum (UHV) MBE growth chamber. To initiate the growth the wafer, was first heated, and a thin Sr layer deposited to assist native oxide desorption. This led to a slight (3 x 6) surface reconstruction when the substrate was cooled down to 500°C. At this temperature the Sr interfacial layer was completed, achieving a ½ monolayer that forms an oxidation barrier between Si and SrTiO$_3$. This was confirmed by a (2 x 1) surface reconstruction on the RHEED pattern. After native oxide removal, direct epitaxy of the first 3 nm of SrTiO$_3$, with [100] SrTiO$_3$ (001) // [110] Si (001) (i.e. 45° rotation of the SrTiO$_3$ crystal with respect to Si) was performed in molecular oxygen at 350°C. After this, growth was paused to switch to atomic oxygen in the growth chamber and to increase the substrate temperature to 550°C. Under these conditions, the remaining SrTiO$_3$ epitaxy was completed at a growth rate of approximately 1 nm/min. Complementary information can be found in Supplementary 1.

### Fabrication

**Wafer Bonding Process:** After growth, the wafer was annealed in oxygen at 850°C for 30 min to reduce the oxygen vacancy concentration in the SrTiO$_3$ layer caused by the reduced pressure environment during growth. Next SiO$_2$ was deposited on top of SrTiO$_3$ by chemical vapor deposition (CVD) to enable subsequent wafer-to-wafer bonding with a SiO$_2$ on Si wafer. The top Si was then removed by grinding followed by a TMAH etch to end up with 105 nm SrTiO$_3$ on top of 2 µm SiO$_2$ on a Si substrate. Finally, an additional annealing in oxygen at 850°C for 30 min was performed.

**Interdigitated finger capacitor (IDC) fabrication:** Al electrodes forming the interdigitated capacitors for RF measurements were patterned using a bilayer liftoff process. Photolithography was used to expose the resist and following the development, 466 nm of Al was deposited using electron beam evaporation. After liftoff the devices were diced into small dies containing the different IDC geometries.

**Electro-optic device:** The process is illustrated in Extended Data Figure 1. Post wafer bonding, ~ 30 nm of chromium and ~ 140 nm of platinum were sputtered on the diced coupons using a bilayer liftoff process and electron beam lithography (EBL) for patterning. Following the liftoff ~ 480 nm polymethyl methacrylate (PMMA) was spin coated while coupling gratings and waveguides with ~ 1.7 µm width were patterned using EBL. The meandered electrodes are roughly 15.37 mm long with a 10 µm electrode gap, 400 µm bending radius, 1.7 µm waveguide width and 200 µm unbalance in the two MZI arms [Supplementary 4.1].

### RF measurement

Permittivity of thin-film SrTiO$_3$ was extracted by performing RF reflectometry analysis on interdigit finger capacitors (IDCs). An Attocube attoDRY800 closed cycle cryostat was used to lower the temperature of the IDCs to below 5 K. The capacitors formed by the interdigit electrodes, see

Extended Data Figure 2 a), were electrically contacted using a GS probe at frequencies from 9 kHz to 14 GHz using a Keysight P5003B vector network analyzer (VNA). An excitation power of 0 dBm was applied, resulting in a probing field below 0.15 Vµm$^{-1}$. This small voltage can be considered a small signal perturbation with respect to the coercive field of SrTiO3. The extraction of permittivity was limited to the range from 100 kHz to 1 GHz due to high-impedance behavior of the capacitor at low frequencies and the impact of resonances on the measurement at high frequencies. The aforementioned resonances are not material resonances; they arise from an increasing parasitic inductive impedance contribution at high frequencies, leading to series LC resonance behavior. Verification of the parasitic nature is obtained through changes to device geometry which controllably reduce the capacitance and correspondingly shift the resonance position up, as can be seen in Extended Data Figure 3 b).

All RF permittivity measurements reported here are at 0 degrees assuming a multidomain [Supplementary Figure 7 c)] mixed c- and a-axis structure in the thin-film plane, leading to an "effective" permittivity. A selection of IDC devices with varying finger lengths from 100 μm to 800 μm was used, the finger gap ($w_{\text{gap}}$) and width ($w_{\text{Finger}}$) were kept constant at 4 μm.

The impedance ($Z$), whose magnitude is seen in Extended Data Figure 3 a), was calculated based on the amplitude and phase of the reflected signal ($S_{11}$) via:

$$Z = 50\,\Omega \cdot \frac{1 + S_{11}}{1 - S_{11}}. \tag{1}$$

The phase of the impedance measurement showed a pure capacitive (-90°) behavior in Extended Data Figure 3 b). Thus, the impedance was mapped to a single total device capacitance ($C$) via:

$$\frac{1}{jZ\omega} = C. \tag{2}$$

We used COMSOL multiphysics simulation of the cross section in

Extended Data Figure 2 b) to match the measured capacitance to an effective permittivity of the SrTiO3. Simulations show that SiO2 and Si contributions can be neglected in $C$ due to the very low permittivity, in strong contrast to that of SrTiO3. To emphasize the negligibility of the SiO2 (and implicitly the Si substrate) contribution we outline a first order estimation below. The parallel capacitance contribution of SrTiO3 ($C_{\text{SrTiO}_3}$) can be approximated by a parallel plate capacitor due to the large relative permittivity ($\varepsilon_r$) where $h_{\text{SrTiO}_3}$ is the height of the SrTiO3 layer and $\#_{Fingers}$ is the number of IDC fingers:

$$C_{\text{SrTiO}_3} = \varepsilon_{r,\text{SrTiO}_3} \cdot \varepsilon_0 \cdot \frac{(L \cdot \#_{\text{Fingers}} \cdot h_{\text{SrTiO}_3})}{w_{\text{gap}}}. \tag{3}$$

Similarly, the perpendicular SiO2 capacitance ($C_{\text{SiO}_2}$) can also be approximated by a parallel plate capacitor:

$$C_{\text{SiO}_2} = \varepsilon_{r,\text{Si}} \cdot \varepsilon_0 \cdot \frac{w_{\text{Finger}} \cdot L \cdot \#_{\text{Fingers}}}{h_{SiO2}}. \tag{4}$$

For the given geometry their ratio is:

$$\frac{C_{\text{SrTiO}_3}}{C_{\text{SiO}_2}} = \frac{\varepsilon_{r,\text{SrTiO}_3}}{\varepsilon_{r,\text{Si}}} \cdot \frac{h_{\text{SrTiO}_3} \cdot h_{\text{SiO}_2}}{w_{\text{gap}} \cdot w_{\text{Finger}}} \approx \frac{\varepsilon_{r,\text{SrTiO}_3}}{\varepsilon_{r,\text{Si}}} \cdot \frac{1}{100}. \tag{5}$$

Consequently, the thin-film capacitance contributes approximately 10 times more than the buried oxide capacitance. In the above arguments it was shown that the measured device capacitance is dominated by the SrTiO3 layer.

The frequency dependent permittivity response indicates there are no significant material resonances within the measured range. Such material resonance contributions would result in a significantly reduced permittivity at higher frequency and thus limit the electro-optic bandwidth potential of $SrTiO_3$. The primary transverse optical phonon mode ($TO_1$) responsible for the high permittivity of $SrTiO_3$ is only expected to occur as high as 2.75 THz [28]. Therefore, the possibility for even higher operating bandwidths before encountering large losses due to intrinsic (material) limitations associated with the $TO_1$ phononic mode exist. Moreover, extrinsic loss mechanisms arising from material defects are expected to play a more dominant role in this respect.

### MZI time domain signal & Pockels coefficient extraction

The electrically modulated time signal of the optical power is governed by the MZI's transfer function. In our design the grating couplers are too close to the splitter and higher order modes leaking to the splitter resulted in a limited extinction ratio due to asymmetric power splitting. For an asymmetric splitting ratio $k$ the MZI transfer function is given by [29]:

$$P(t) \propto |E_{out}|^2 = P_0 \left((1-k)^2 + k^2 + 2k(1-k)\cos\left(\frac{2\pi}{\lambda}\cdot \Delta n_{eff} \cdot L + \theta\right)\right). \tag{6}$$

Here $\theta$ corresponds to the MZI's zero voltage operation point and is given by the 200 μm unbalance of the two MZI arms. The laser wavelength is $\lambda$, $L$ is the electrode length and $\Delta n_{eff}$ is the voltage induced effective refractive index difference [Supplementary 4.2]. The splitting ratio ($k$) was extracted from the time signal by solving:

$$\frac{P_{min}}{P_{max}} = 1 - 4k + 4k^2. \tag{7}$$

The optical power can be transformed to a phase via:

$$\Psi(t) = \cos^{-1}\left(\frac{\left(\frac{P(t)}{P_0}\right) - (1-k)^2 - k^2}{2k(1-k)}\right) = \theta + \frac{2\pi}{\lambda}\cdot \Delta n_{eff} \cdot L, \tag{8}$$

resulting in Extended Data Figure 4. The phase depicted is wrapped by the arccosine function and needs to be unwrapped for every turning maximum/minimum. Normally, the phase should oscillate between 0 and $\pi$, however, the 10 kHz bandwidth of the photodiode in combination with the strong overmodulation (~ $30\pi$ phase change) caused a low-pass behavior that prevents the full oscillations from 0 to $\pi$. These low-pass characteristic was compensated in the post processing. After unwrapping and subtraction of the operation point, $\theta$ results in a continuous phase change curve vs applied field, as shown in the main text.

The change in effective refractive index can then be extracted using [30]:

$$\Delta n_{eff} = \Gamma \cdot \frac{\Delta n_{mat}}{n_0} \cdot n_g = \Gamma \cdot \left(\frac{1}{2}\cdot n_0^2 \cdot r_{eff} \cdot v \cdot E_{AC} + R \cdot E_{AC}^2\right) \cdot n_g, \tag{9}$$

Where $E_{AC}$ is the electric field induced by the applied sinusoidal voltage. For detailed information see Supplementary 4.2.

# Extended Data Figures

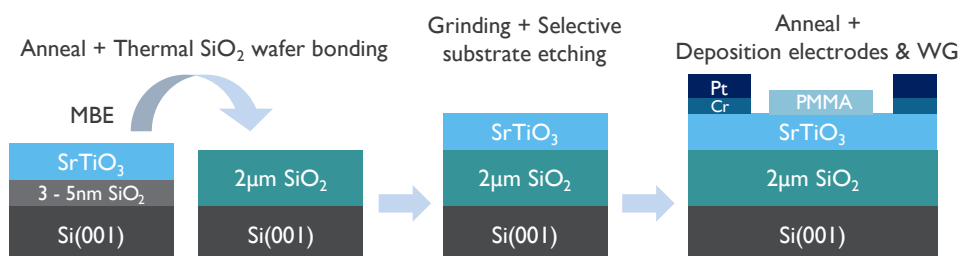

Extended Data Figure 1) Illustration of the electro-optic (EO) device fabrication. After SrTiO$_3$ epitaxy on Si (001), the wafer was annealed in oxygen and thermally wafer-bonded onto a SiO$_2$ box to optically isolate the SrTiO$_3$ film from the silicon substrate. Next, the original Si substrate was removed by grinding and selective substrate TMAH etching. Finally, the film was annealed once more after which the Cr/Pt electrodes and PMMA waveguides were deposited.

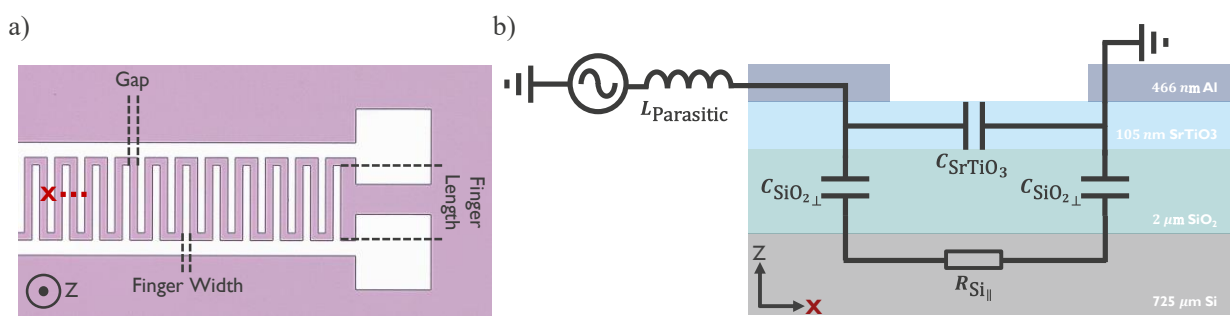

Extended Data Figure 2 a) Out-of-plane optical microscope image of an IDC capacitor. Geometrical features are annotated, red dashed line shows the representative location of the cross section seen in b). b) Simplified equivalent circuit model considerations including in/out-of-plane capacitances and resistances for the SrTiO$_3$/SiO$_2$/Si IDC.

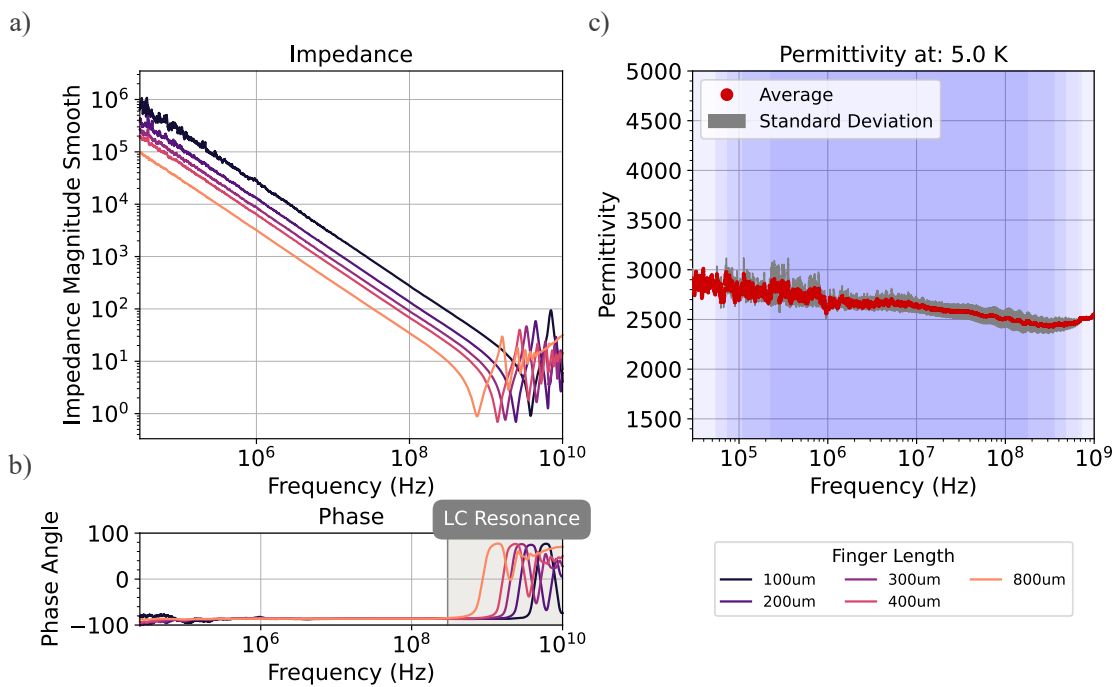

Extended Data Figure 3 a) Impedance magnitude and b) phase angle of a subset of capacitors, 4 $\mu$m gap and varying finger length (denoted by color in legend), used to extract c) permittivity at 5 K from 30 kHz to 1 GHz. The reduced standard deviation in the highest and lowest frequency regimes arises from the measurement of fewer devices, limited by the frequency range of a particular capacitor. Lightest blue shaded regions represent data obtained with a single device while increasingly darker shaded regions represent an addition of a single device, the darkest region represents the average permittivity of 5 devices.

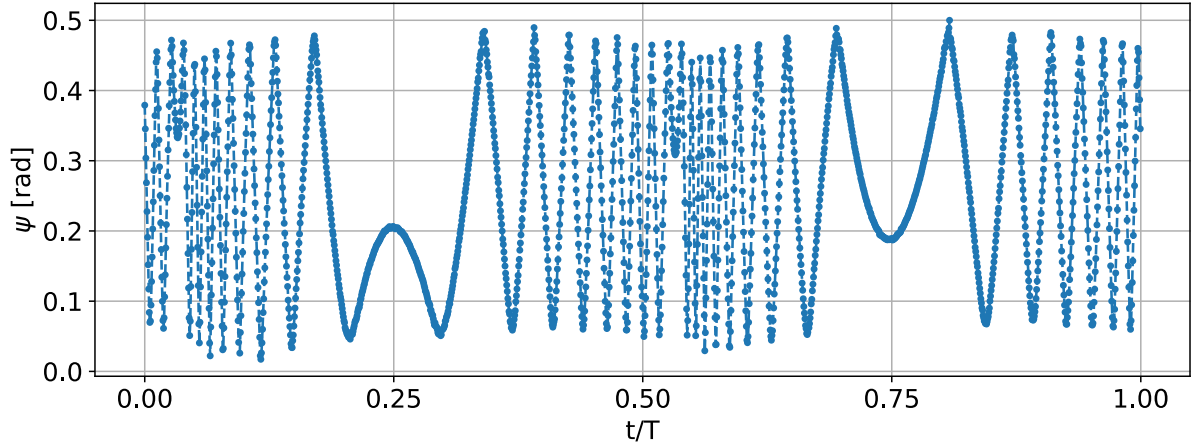

Extended Data Figure 4) Phase vs one electrical modulation period.

## Acknowledgements

We thank the electron beam lithography team for their expert assistance with sample fabrication, the RBS team for the characterization of our material, Sean McMitchell for the fruitful crystallographic discussions and Nathaniel Kinsey for providing valuable feedback.

This research was funded by the Branco-Weiss Society and the European Research Council (ERC) under the European Union's Horizon 2020 research and innovation program, grant number 864483 NOTICE and grant number 101042414 Q-AMP.

## End Notes

**Author contributions** A.U. designed electro-optic and cutback structures, fabricated and measured electro-optic sample. K.B designed, fabricated and measured cryogenic to RT RF devices. A.B epitaxially grew $SrTiO_3$, performed wafer characterization, fabricated optical cutback structures and performed room temperature optical loss measurements. M.D. assisted cryogenic rf measurements. C.S assisted electro-optic measurements. Y.H helped with the electron-beam-lithography exposure of samples. S.S.S assisted waferbonding flow in fabrication line. M.B supported epitaxial growth and performed RSM characterization and analysis. P.F performed TEM and EDS characterization and analysis. S.S and T.N. performed Raman characterization and analysis. K.V.G, B.K, K.D.G, C.M and C.H conceived and supervised the project. A.U., K.B., A.B., K.D.G and C.H. prepared the manuscript, with input from all authors.
**Competing interest** The authors declare no competing interests.
**Supplementary information** Supplementary Information is available for this paper.
**Correspondence and requests for materials** Correspondence and requests for materials should be addressed to Christian Haffner.

# Supplementary Information
# Engineering high Pockels coefficients in thin-film strontium titanate for cryogenic quantum electro-optic applications

Anja Ulrich[1,2*], Kamal Brahim[1,3*], Andries Boelen[1,4*], Michiel Debaets[1,2], Conglin Sun[1,5], Yishu Huang[1,2], Sandeep Seema Saseendran[1], Marina Baryshnikova[1], Paola Favia[1], Thomas Nuytten[1], Stefanie Sergeant[1], Kasper Van Gasse[1,2], Bart Kuyken[1,2], Kristiaan De Greve[1,3], Clement Merckling[1,4], Christian Haffner[1†]
* contributing equally, † corresponding author

1) Imec, B-3001 Leuven, Belgium
2) Department of Information Technology (INTEC), Photonics Research Group, Ghent University, B-9052 Ghent, Belgium
3) Department of Electrical Engineering (ESAT), KU Leuven, B-3001 Leuven, Belgium
4) Department of Materials Engineering (MTM), KU Leuven, B-3001 Leuven, Belgium
5) Department of Physics and Astronomy, KU Leuven, B-3001 Leuven, Belgium

## Supplementary 1. Growth

The quality of the epitaxial layer is key for implementing $SrTiO_3$ as an active material in quantum applications. In our previous work, we demonstrated the critical role of stoichiometry and $SrTiO_3$ thickness on the structural, optical and electrical properties [1]. $SrTiO_3$ films which are slightly off in stoichiometry still resulted in a good crystalline quality, which is essential for the film in this work (Sr/Ti = 0.96 ± 0.03). Supplementary Figure 1 shows the dependence of cationic stoichiometry (Sr/Ti) and $SrTiO_3$ (002) ω-scan FWHM on the effective permittivity $\varepsilon_r$ at cryogenic temperatures for four different $SrTiO_3$ films. In an ω-scan, the XRD detector is fixed at an unchanging angle while the sample is tilted slightly. To maximize the permittivity, achieving a minimal FWHM is crucial, which can be attained by approaching ideal stoichiometry and increasing the $SrTiO_3$ thickness.

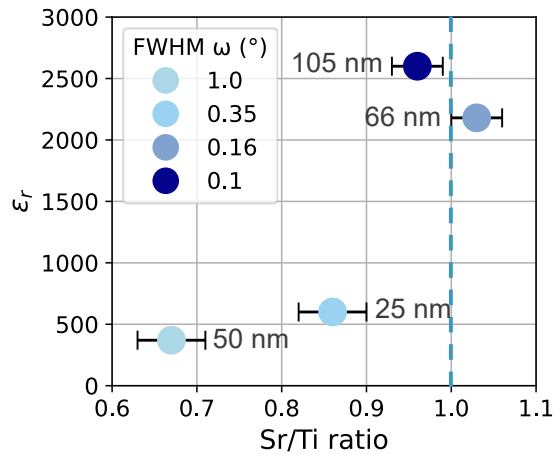

Supplementary Figure 1) Cryogenic relative permittivity ($\varepsilon_r$) at as a function of cationic stoichiometry (Sr/Ti ratio) for four $SrTiO_3$ films with varying thickness and FWHM of the $SrTiO_3$ (002) ω-scan. Sr/Ti near 1 and greater thickness reduce the FWHM and enhance the permittivity.

## Supplementary 2. Theoretical thermal strain calculations at 4 K

Besides the ferroelectric behaviour of our SrTiO$_3$ due to stoichiometry, strain can also elevate the Curie temperature (T$_C$) and contribute to a ferroelectric phase transition at cryogenic temperatures. This dependency can be seen on the phase diagram in Supplementary Figure 2, which illustrates the behavior of T$_C$ as a function of in-plane strain [2]. Room-temperature reciprocal space mapping (RSM) indicates a fully relaxed SrTiO$_3$ layer with unit cell equal to bulk, see Supplementary Figure 3. The relaxation in epitaxial films can be facilitated by dislocation gliding during thermal treatment. Nevertheless, strain can be induced at cryogenic temperatures due to the difference in thermal expansion coefficients (CTE) between film and substrate (CTE$_{SrTiO3}$ = 8.8 · 10$^{-6}$ K$^{-1}$, CTE$_{Si}$ = 2.6 · 10$^{-6}$ K$^{-1}$) [3]. Because the Si substrate (725 μm) is much thicker than the SiO$_2$ layer (2 μm), the SiO$_2$ layer, and consequently the SrTiO$_3$ film, will contract following the Si substrate upon cooling. Assuming our SrTiO$_3$ film to be isotropic, the induced biaxial thermal strain can be calculated by [4]:

$$Strain_{\text{thermal,SrTiO}_3} = \left(CTE_{\text{SrTiO}_3} - CTE_{\text{Si}}\right) \cdot \Delta T . \tag{10}$$

At 4 K this results in a tensile strain of 0.18%, which would be the upper limit of the induced strain at cryogenic temperatures for our sample. 0.18% strain (blue cross) corresponds to 50 K < T$_C$ < 110 K [2] aligning with the permittivity measurements and reported T$_C$ in the main text.

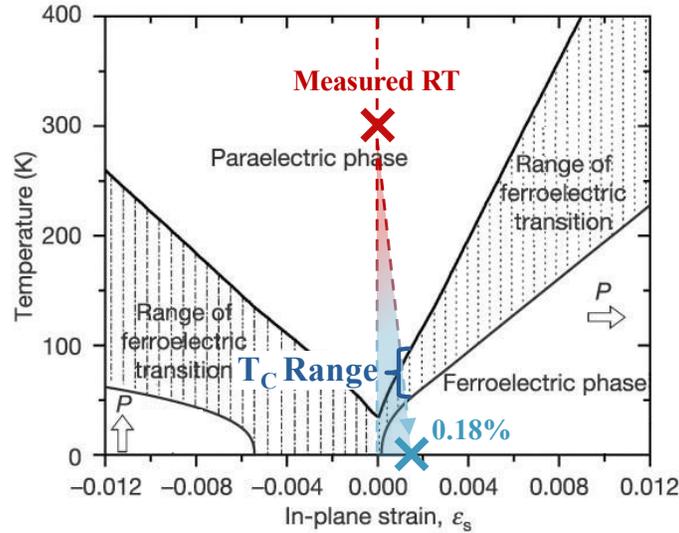

Supplementary Figure 2) Phase diagram of bulk SrTiO$_3$ illustrating the relationship between in-plane strain and Curie temperature T$_C$. At room-temperature, reciprocal space mapping (RSM) indicates a fully relaxed SrTiO$_3$ film, which is consequently in a paraelectric phase (red cross). Relaxed SrTiO$_3$ at 4 K is quantum paraelectric, our upper estimation of 0.18% strain due to the difference in thermal expansion coefficients (CTE) would induce a ferroelectric phase transition (blue cross) with 50 K < T$_C$ < 110 K. Figure adapted from [2].

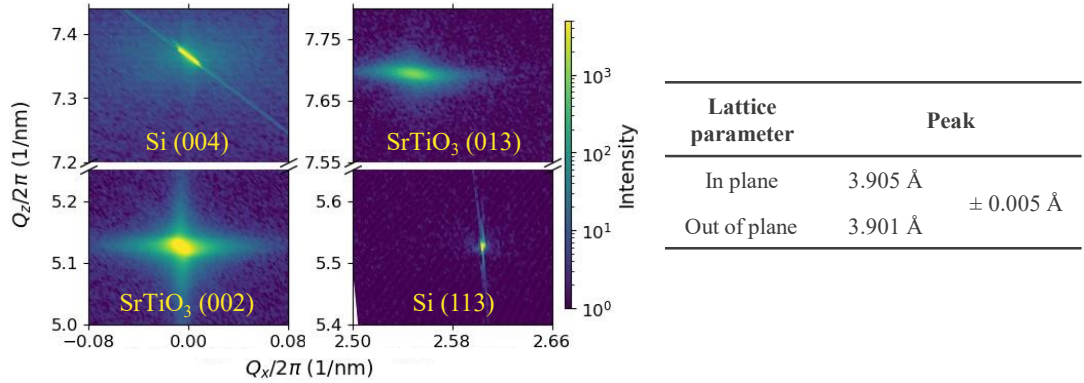

Supplementary Figure 3) Room-temperature XRD reciprocal space map (RSM) shows the symmetric Si (004) and $SrTiO_3$ (002), and asymmetric Si (113) and $SrTiO_3$ (013) diffraction peaks. The peak positions indicate a fully relaxed cubic $SrTiO_3$ unit cell with lattice parameter equal to bulk ($a_{bulk}$ = 3.905 Å).

## Supplementary 3.  Relaxor ferroelectricity

Temperature-dependent Raman spectroscopy shows the gradual appearance of the Slater and Last phononic mode at 5.2 THz upon cooling. Supplementary Figure 4 (right) illustrates the fitted peak intensity by an asymmetric profile as a function of temperature. The peak at 5.2 THz starts to appear below 200 K. The precise temperature of this phononic modes' appearance is independent of whether the sample is cooled down (blue) or warmed up (red). The presence of the Raman signal is a clear indicator of ferroelectricity, as this soft mode is only Raman active under non-centrosymmetric structure conditions. Literature links the gradual appearance of the signal to the relaxor behavior coming from polar nanoregions [5–7].

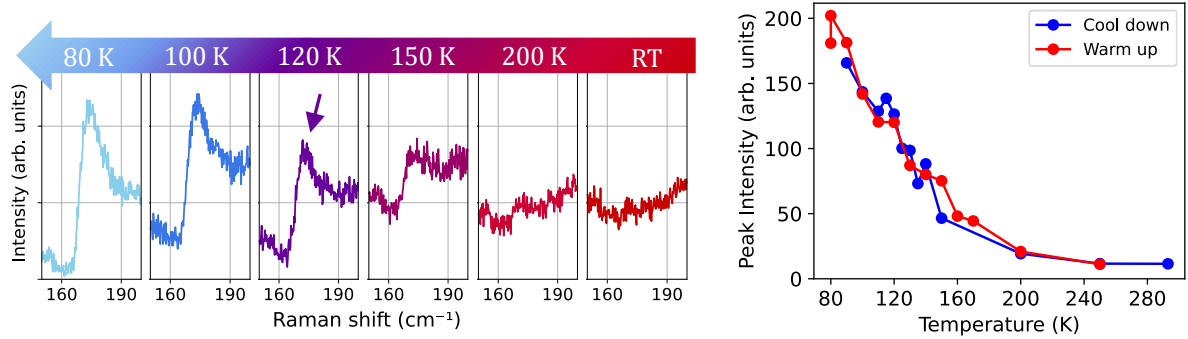

Supplementary Figure 4) Appearance of the ferroelectric $SrTiO_3$ phononic mode when cooling down (left). Fitted peak intensity of the mode at 5.2 THz (~ 173 cm$^{-1}$) as a function of temperature shows the gradual appearance of the Raman signal below 200 K. This behavior is independent whether cooling down or warming up and indicates a relaxor ferroelectric phase (right).

## Supplementary 4.  Electro-optic measurements

### Supplementary 4.1. Setup & Device

The MZI device is cooled down below 5 K with the same cryostat as the RF setup. However, in and outcoupling via two fibers required us to remove the RF probe shown in Supplementary Figure 5 a). The sample is silver pasted onto the sample holder and the gratings of the sample are interfaced using angled fibers on piezo stages, which can be controlled individually to change the grating coupler coupling efficiency. A tunable Keysight N7778C laser at around 1470 nm is used to send in an optical signal. The electrodes are wirebonded with Al wires to a PCB which is connected to either the Keysight 33512B arbitrary waveform generator (AWG) or the Keysight B2910BL precision source measure unit (SMU). The former can supply up to 20 MHz and 10 V peak sinusoidal drive signal while the latter one can be programmed to output a sinusoidal signal up to 1 kHz with up to 40 V peak voltage. The electrically modulated optical signal at the optical output is detected using a Keysight N7744C photodetector (PD) which has a 10 kHz 3 dB cutoff frequency.

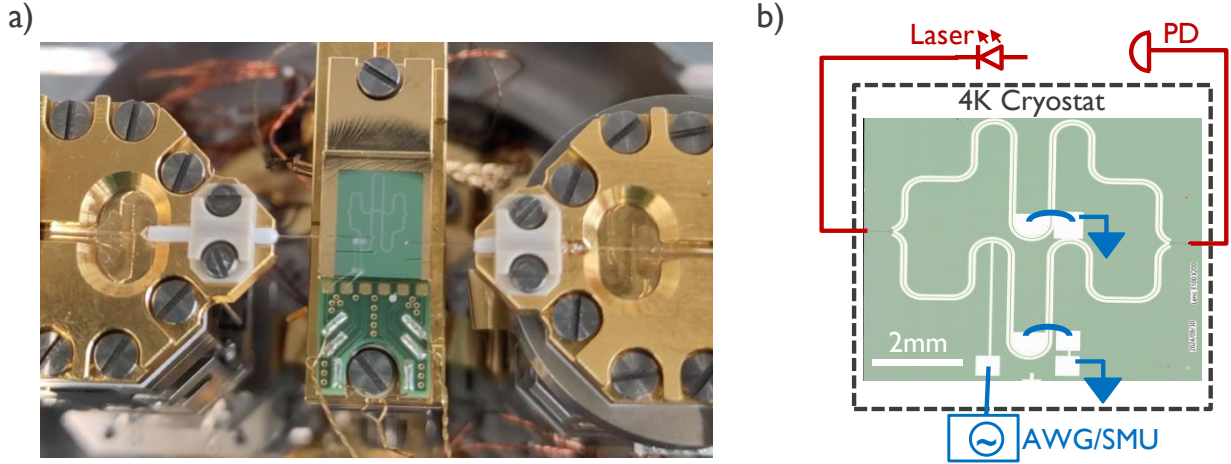

Supplementary Figure 5 a) Electro-optic setup. The fibers are mounted with fiber holders onto piezo stages to the left and right of the sample in the center. Angled fibers are used. The sample is silver pasted onto a sample holder and wirebonded to a PCB. The PCB low frequency wires are then connected to the breakout box at the bottom of the cryostat, connecting to the room temperature electronics. b) Setup schematics with laser, photodiode (PD), arbitrary waveform generator (AWG), precision source measure unit (SMU) and zoom in of the MZI sample.

The sample holder limits the sample size to below 1 cm$^2$, as shown in Supplementary Figure 5 a). Therefore, meandered electrodes have been used instead of straight ones to increase the interaction length and the modulated output signal. The electrodes are roughly 15.37 mm long with a 10 μm electrode gap, 400 μm bending radius, 1.7 μm waveguide width and 200 μm unbalance in the two MZI arms.

## Supplementary 4.2. Effective Pockels coefficient extraction

The change in effective refractive index is given by [8]:

$$\Delta n_{\text{eff}} = \Gamma \cdot \frac{\Delta n_{\text{mat}}}{n_0} \cdot n_{\text{g}} = \Gamma \cdot \left(\frac{1}{2} \cdot n_0^2 \cdot r_{\text{eff}} \cdot \upsilon \cdot E_{\text{AC}} + R \cdot E_{\text{AC}}^2 \right) \cdot n_{\text{g}}, \tag{11}$$

$E_{\text{AC}}$ is the electric field induced by the applied sinusoidal voltage. The data shown in the main figure were obtained by a 20 Hz and 16 V amplitude dropping over 10 μm electrode spacing. The change of refractive index of the optical mode is determined by the material's refractive index change coming from the linear electro-optic Pockels coefficient $r_{\text{eff}}$ and quadratic electro-optic Kerr coefficient $R$. This change scales with the light confinement in the SrTiO$_3$ ($\Gamma$), the poling fraction $\upsilon$ and with the group refractive index $n_{\text{g}}$. With the above definition $\Gamma$ corresponds to the energy confinement factor which is [9]:

$$\Gamma = \frac{\int_{A_{\text{SrTiO}_3}} \varepsilon_r |\boldsymbol{E}|^2 dA}{\int_A \varepsilon_r |\boldsymbol{E}|^2 dA} = 43\%, \tag{12}$$

where the numerator expression is integrated over the SrTiO$_3$ area and the denominator over the overall area of the optical mode. The poling function $v$ is given by [10]:

$$v = \frac{\sinh(U \cdot (E \pm E_C))}{U/\chi - 1 + \cosh(U \cdot (E \pm E_C))},\tag{13}$$

where $U$ is the equivalent field, $E_C$ the coercive field and the linear electric susceptibility $\chi = \varepsilon_r - 1$.

From the unwrapped phase and equation (11) $r_{\text{eff}}$ can be extracted as,

$$r_{\text{eff}} = \frac{d\Delta\psi}{dE_{AC}} \cdot \frac{\lambda}{2\pi \cdot L} = \left(\frac{1}{2} \cdot n_0^2 \cdot \left(r(\varepsilon_r) \cdot \frac{dv}{dE_{AC}} \cdot E_{AC} + r(\varepsilon_r) \cdot v\right) + R \cdot E_{AC}\right) \cdot \Gamma \cdot n_g,\tag{14}$$

which is shown in Supplementary Figure 6 a). The red curve shows a nonlinear decaying trend for negative voltages in the down sweep direction and similarly for the blue curve in the positive voltage up sweep direction. Instead, the curve should flatten out as one would expect from the fully poled case with poling function being equal to one as seen in (b) or a linear decay due to a strong Kerr coefficient as seen in (c). It was not possible to find a good physical fit using equation (14). The origin of the previously mentioned decays is attributed to the Pockels coefficients direct relation to the permittivity $r(\varepsilon_r)$, which falls with increasing bias magnitude [11]. This dependence is not included in equation (14) and requires further investigations.

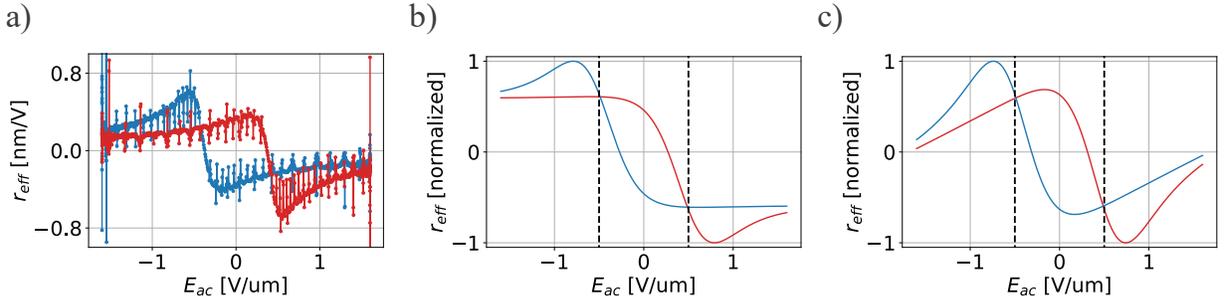

Supplementary Figure 6 a) Effective Pockels coefficient for up (red) and down (blue) electric field sweep direction. b) Normalized example for $r_{\text{eff}}$ with negligible Kerr coefficient showing a flattening out for field values greater than the coercive field (black dashed vertical line). c) Case for non-negligible Kerr coefficient with same values as in b) showing a linear contribution counteracting $r_{\text{eff}}$.

Due to space constraints of the setup the Pockels value given by equation (14) needs to be corrected for the effective contribution of different crystal directions along the waveguide arising from the meandered layout.

According to theoretical studies [12,13], perturbated SrTiO$_3$ is predicted to be in a 4mm or mm2 point group, the former being the point group of BaTiO$_3$. Both point groups have the off-diagonal $r_{51}$ element in plane for an a-axis dominated film. For complex oxides like BaTiO$_3$ the permittivity is typically lower for the c-axis due to a stronger confinement of the Ti atom in the ferroelectric potential wells, which leads to a reduced polarizability. Therefore, it can be expected that the $r_{33}$ and $r_{13}$ coefficients with the RF field along the c-axis of the crystal are lower compared to $r_{42}$ with the RF field along the a-axis (Miller's rule).

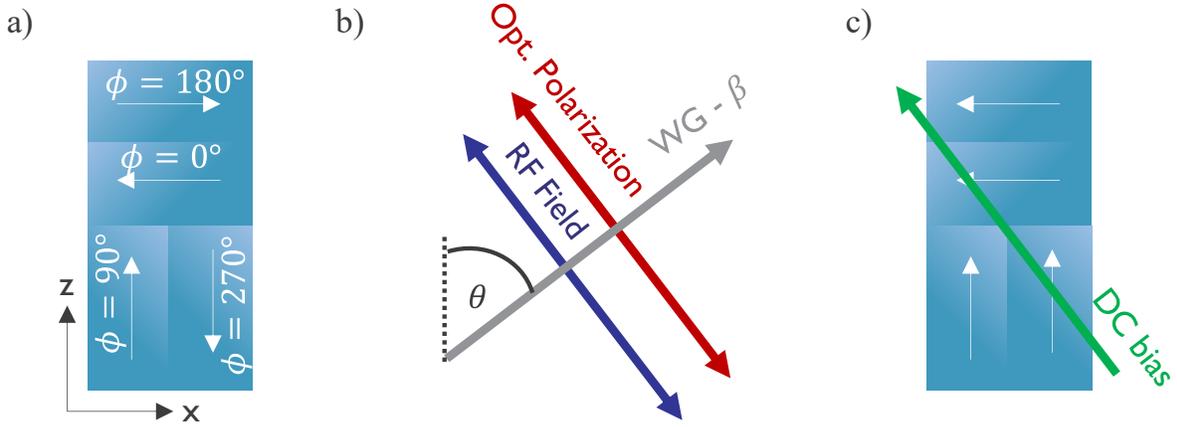

Supplementary Figure 7 a) Schematics of a top view of a-axis film with domains like $BaTiO_3$ or possibly $SrTiO_3$. The white arrow indicates the direction of the spontaneous polarization which is random. For growth reasons, the domains stand perpendicular to each other. b) Illustration of the waveguide orientation (grey) with the optical polarization (red) and the RF field (blue). The electrodes and waveguide run in parallel, aligned at an angle $\theta$ with respect to the crystal domains. c) Schematics of the effect of applying a DC bias and flipping domains resulting in a macroscopic nonzero electro-optic response.

Preliminary measurement along the $r_{33}$ direction confirm a reduced Pockels strength compared to the 45° case. Thus, in the following we neglect the $r_{33}$ contribution and can define the effective Pockels coefficient as [14]:

$$r_{\text{eff}}(\theta) = \sum_\varphi \sigma_\varphi \cdot \left( \cos^2(\theta) \cdot \sin(\theta) \cdot (r_{13} + 2r_{51}) \right), \tag{15}$$

where $\theta$ is the device angle with respect to the z-axis and crystalline domains, see Supplementary Figure 7 a) & b). The domain direction in a) is given by $\phi$ equal to $0°, 90°, 180°$ or $270°$ and corresponds to the optical propagation direction $\theta$ where the RF field points along the domains. Applying a bias significantly stronger than the coercive field will result in flipping of domains as shown in c) and $v = \pm 1$. The light travelling in direction $\theta$ with the RF and optical polarization (TE mode) perpendicular to $\theta$ experiences different domains orientations. This is reflected in the sum sign and the averaging factor $\sigma_\varphi$ in equation (15) with condition $\sum_\varphi \sigma_\varphi = 1$. For the simplified fully poled case there will be only two contributing domain directions, i.e. $\sigma_\varphi = 0.5$, namely $\theta + 0°$ and $\theta + 90°$ and equation (15) can be rewritten to:

$$r_{\text{eff}}(\theta) = 0.5 \cdot \left( (\cos\theta \sin^2\theta + \sin\theta \cos^2\theta) \cdot \text{sgn}(v_\varphi) \cdot (r_{13} + 2r_{51}) \right). \tag{16}$$

Here the $\cos\theta$ term maps onto the $\theta + 90°$ crystal domain and the $\sin\theta$ term maps onto the $\theta + 0°$ domain. In the equation we included the $\text{sgn}(v_\varphi)$ of the poling function as we propagate along different directions, with the poling function being either plus or minus one, in our device. We use the same definition as for the electric field mapping, i.e. the $\phi = 0°, 90°$ domains have a positive poling polarity and the $\phi = 180°, 270°$ a negative poling polarity.

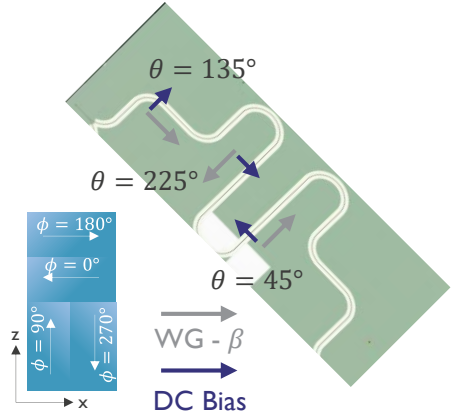

Supplementary Figure 8) Schematics of the 45 degree sample with propagation and DC bias/RF field direction with respect to the z-axis for the different sections of the sample.

For the straight parts of our design $\theta$ equals 45°, 225° or 135°, as shown in Supplementary Figure 8 which equals $|\cos\theta \sin^2\theta| + |\sin\theta \cos^2\theta| = 1/(\sqrt{2})$ for all directions and involves two domains which determine the sign. Neglecting the electrode bends, we can write the effective measured Pockels coefficient of our device as:

$$r_{\text{eff,meas}} = 0.22 \cdot r_{\text{eff}}(\theta = +45°) + 0.23 \cdot r_{\text{eff}}(\theta = +225°) + 0.08 \cdot r_{\text{eff}}(\theta = 135°), \quad (17)$$

where the straight parts have been scaled by the percentage of overall length in this direction. I.e. for 22% of the electrode length light travels in the 45° direction, where the fully poled 0° and 90° domain are experienced equally strong ($\sigma_\varphi = 0.5$), both contributing a factor $1/(2\sqrt{2})$ with a positive polarity. Therefore, $r_{\text{eff}}(\theta = +45°) = 0.5 \cdot (2 \cdot 1/(2\sqrt{2})) \cdot (r_{13} + 2r_{51}) = 1/(2\sqrt{2}) \cdot (r_{13} + 2r_{51})$. The other directions yield:

| $r_{\text{eff}}(\theta)$ | $\sigma_\varphi \cdot (\text{sgn}(v_\varphi) \cdot \cos\theta \sin^2\theta + \text{sgn}(v_\varphi) \cdot \sin\theta \cos^2\theta)$ | $r_{\text{eff}}$ |
|---|---|---|
| $r_{\text{eff}}(\theta = 45°)$ | $0.5 \cdot \left((+1) \cdot \left(\frac{1}{2\sqrt{2}}\right) + (+1) \cdot \left(\frac{1}{2\sqrt{2}}\right)\right)$ | $1/(2\sqrt{2}) \cdot (r_{13} + 2r_{51})$ |
| $r_{\text{eff}}(\theta = 225°)$ | $0.5 \cdot \left((-1) \cdot \left(\frac{-1}{2\sqrt{2}}\right) + (-1) \cdot \left(\frac{-1}{2\sqrt{2}}\right)\right)$ | $1/(2\sqrt{2}) \cdot (r_{13} + 2r_{51})$ |
| $r_{\text{eff}}(\theta = 135°)$ | $0.5 \cdot \left((-1) \cdot \left(\frac{-1}{2\sqrt{2}}\right) + (+1) \cdot \left(\frac{1}{2\sqrt{2}}\right)\right)$ | $1/(2\sqrt{2}) \cdot (r_{13} + 2r_{51})$ |

Supplementary Table 1) Effective Pockels coefficient summarized for the three straight electrode directions given by $\theta$

The same needs to be done for the bends ($\sim 2.5 \cdot 2\pi$) contributing 47% to the overall length. Here we assumed the simplified full poled case as well, which overestimates the bend contribution. Due to symmetry, it is sufficient to calculate the integral for a quarter circle. This gives:

$$r_{\text{eff,quarter bend}} = \frac{1}{\frac{\pi}{2}} \cdot \left(\int_0^{\frac{\pi}{2}} \frac{1}{2} \cdot (\cos\theta \sin^2\theta + \sin\theta \cos^2\theta) d\theta \cdot (r_{13} + 2r_{51})\right) \quad (18)$$
$$= \frac{1}{\pi} \cdot \frac{2}{3} \cdot (r_{13} + 2r_{51}).$$

Thus equation (17) becomes:
$$r_{\text{eff,meas}} = 0.22 \cdot r_{\text{eff}}(\theta = +45°) + 0.23 \cdot r_{\text{eff}}(\theta = +225°) + 0.08 \cdot r_{\text{eff}}(\theta = 135°) \\ + 0.47 \cdot r_{\text{eff,quarter bend}}. \tag{19}$$

Using the calculated expression in Supplementary Table 1 and equation (18) yields:
$$r_{\text{eff,meas}} = \frac{(r_{13} + 2r_{51})}{2\sqrt{2}} \cdot \left(0.22 + 0.23 + 0.08 + 0.47 \cdot \frac{2}{3\pi} \cdot 2\sqrt{2}\right) \approx 0.81 \cdot r_{\text{eff}}. \tag{20}$$

The compensated effective value reported in the main text is:
$$r_{\text{eff}} = \frac{(r_{13} + 2r_{51})}{2\sqrt{2}} = \frac{r_{\text{eff,meas}}}{0.81} \approx 345 \text{ pm/V}, \tag{21}$$

here $r_{\text{eff,meas}} \propto \Delta n_{\text{eff}}/L \approx 280$ pm/V corresponds to the experimentally extracted phase shift (equation (9)) divided by the full length, not accounting for the correction from different directions. By solving equation (9) for a $\pi$ phase shift the half-wave-voltage length product $V_\pi L$ can be calculated:
$$V_\pi L \approx E_{AC} \cdot d \cdot L = \frac{\lambda_0 \cdot d}{\Gamma \cdot n_0^2 \cdot n_g \cdot r_{\text{eff}}} \approx 1.04 \pm 0.08 \text{ Vcm}, \tag{22}$$

where $d = 10$ µm is the electrode distance.

Using $r_{\text{eff}} \approx 345$ pm/V and the zero-bias voltage permittivity $\varepsilon_r \approx 2700$ we can estimate our Miller's coefficient [15]:
$$\delta_{M2} = \frac{r_{\text{eff}} \cdot n_0^4}{2(\varepsilon_r - 1) \cdot (n_0^2 - 1)^2} \approx 0.1 \cdot 10^{-12} \frac{\text{m}}{\text{V}}. \tag{23}$$

### Supplementary 4.3. Small signal modulation

To verify the poling from the chapter before, a strong but slowly varying sinusoidal signal (20 Hz, 1.6 µm/V) combined with a weak and fast varying sinusoidal signal (1 kHz, 0.05 V/µm), close to the limit of our photodetector, was applied as shown in Supplementary Figure 9 a). Therefore equation (8) is changed to:
$$\Psi(t) = \cos^{-1}\left(\frac{\left(\frac{P(t)}{P_0} - (1-k)^2 - k^2\right)}{2k(1-k)}\right) = \theta + \frac{2\pi}{\lambda} \cdot (\Delta n_{\text{eff,20Hz}} + \Delta n_{\text{eff,1kHz}}) \cdot L \tag{24}$$

containing both frequency contributions. To extract the contribution from the 1 kHz signal the maxima and minima of the drive signal without the fast signal can be used for overlapping as shown in b). The subtraction of the 20 kHz contribution can then be fitted and yields $r_{\text{eff,1kHz}} \approx 134 \pm 2$ pm/V. This confirms the value obtained from the poling curve in the main text for maximal poling field of 1.6 V/µm and relies on the bandwidth of $r_{\text{eff}}$ being constant up to 1 kHz.

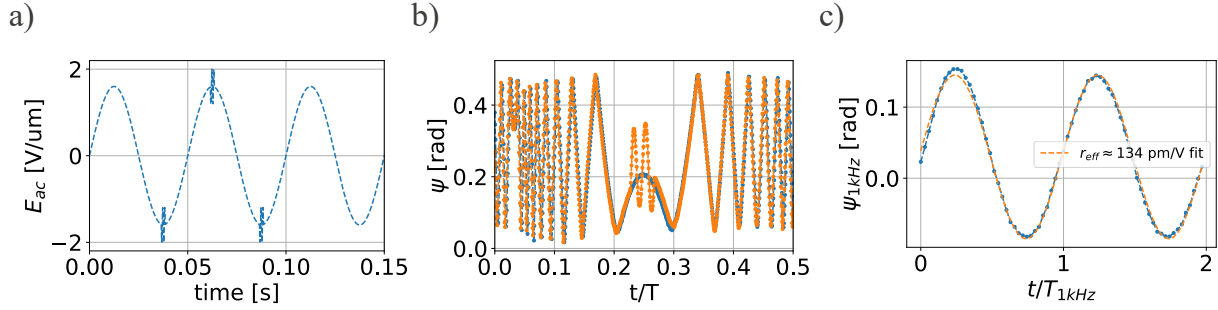

Supplementary Figure 9 a) Drive signal composing of slow 20 Hz signal with 1.6 V/µm amplitude and fast 1 kHz 0.05 V/µm signal. b) Power signal with and without fast oscillation overlapped and transformed to phase as given by equation (24). c) Subtracted phase signal containing only the 1 kHz contribution with fit.

### Supplementary 4.4. Bandwidth

Due to the low optical output power of - 50 dBm of our device and the limited bandwidth of the used PD, the time signal cannot be resolved anymore for frequencies beyond 1 kHz. Therefore, a time averaging method was used to measure the electro-optic frequency response beyond 1 kHz. As shown in the inset in Supplementary Figure 10 a), the laser is set to a constructive interference point ("on"-point) of the MZI and a weak ac signal of 0.1 V/µm is applied to minimize the effect of any third-order nonlinear contributions. The time averaged power at the PD is reduced due to the symmetric oscillations around the on-point with an averaging time $t_{\text{avg}} \gg 1/f_{\text{AC}}$:

$$P_{\text{on}} - P(t) = P_{\text{on}} - \frac{P_{\text{on}}}{t_{\text{avg}}} \int_0^{t_{\text{avg}}} \left(\frac{1}{2} + \frac{1}{2} \cdot \cos\left(\frac{2\pi}{\lambda} \cdot \Delta n_{\text{eff}}(t) \cdot L\right)\right) dt \geq 0. \quad (25)$$

The reduction is directly proportional to the strength of the effective Pockels coefficient at the different frequencies as given by equation (25) with $\Delta n_{\text{eff}} \propto r_{\text{eff}} \cdot v \cdot E_{\text{AC}}(t)$. The optical power difference is unaffected by an increase in ac modulation frequency, indicating a constant effective Pockels coefficient up to 1 MHz. The bandwidth curve in Supplementary Figure 10 a) shows a resonance at 5 MHz (vertical dashed line). The source of the 5 MHz resonance was identified by characterizing the low frequency lines without a sample connected. An S11 reflection measurement as shown in b) shows that the system (without sample) has resonances appearing from 1 MHz onwards; therefore, limiting higher speed measurements and clarifying the origin of the resonance seen in our bandwidth data. The reduced response time for frequencies below 50 Hz, preventing DC bias experiments, is believed to be due to screening.

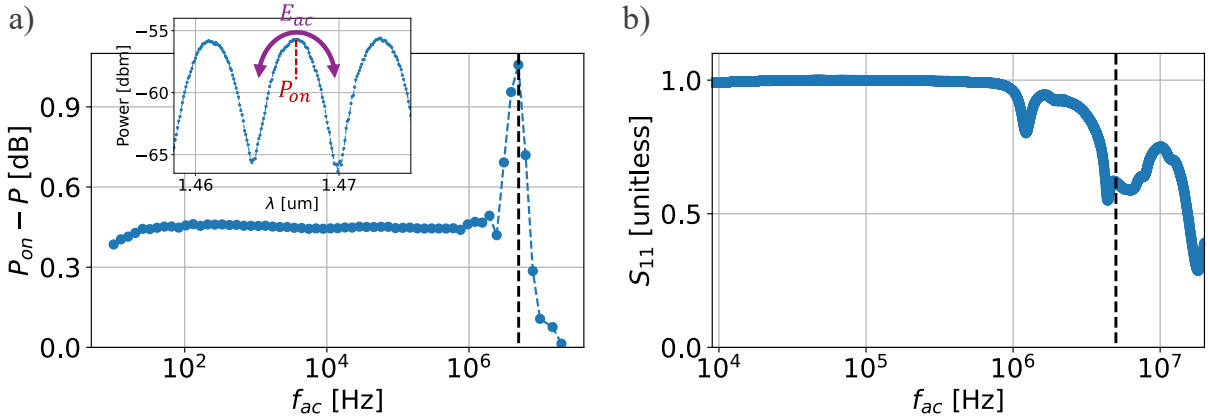

Supplementary Figure 10 a) Bandwidth measurement as described in main text up to 20 MHz. The Data shows a resonance peak at 5 MHz (vertical dashed line). b) S11 data of the low frequency lines of our setup without a sample connected (open). The S11 data shows that after 1 MHz the system itself seems to have some intrinsic resonances.

## Supplementary 4.5. DC screening

Applying a DC bias to the electrodes initially induces a change in the optical output signal. However, this is followed by a several second long decay of the optical power towards the no-bias baseline of the MZI. This screening behaviour is shown in Supplementary Figure 11 a) for different bias voltages. The measurement indicates a faster decay for higher voltages.

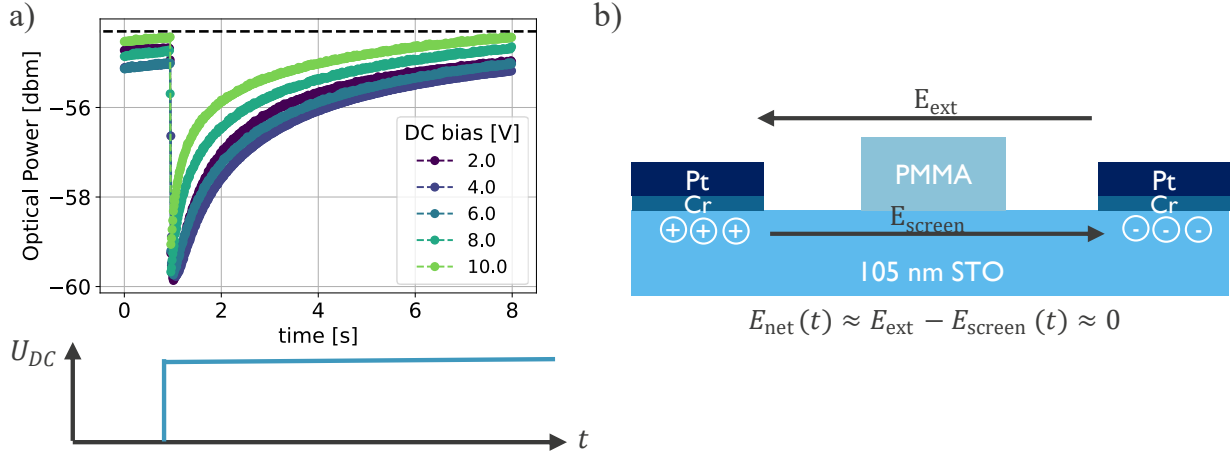

Supplementary Figure 11 a) Optical power vs time for different applied DC bias. The DC bias is applied at around t = 1 s and kept constant, as indicated by the plot at the bottom. The measurement shows a clear decay with applied DC bias back to the baseline (indicated by vertical dashed line). Stronger biases decay faster than low biases. b) The hypothesis for the decay of the nonlinear response is the buildup of a screening field within the material, caused by charge carriers like oxygen vacancies.

For memristors it is shown that bulk $SrTiO_3$ with the same metals as in our device can create or absorb oxygen vacancies depending which voltage polarity is applied [16]. These vacancies could build up an internal screening field as shown in Supplementary Figure 11 b) which could lead to a net electric field of zero within the nonlinear material and therefore to a vanishing effective Pockels coefficient. As the room temperature diffusion coefficient of oxygen vacancies $D \approx$ 1e-14 cm²/s [17] is low and should be even lower for cryogenic temperatures, a diffusion of generated vacancies across the electrode gap of 10 µm can be excluded.

The TEM and energy-dispersive X-ray spectroscopy (EDS) results of the device at two different positions are shown in Supplementary Figure 12 a). The EDS under the electrodes in b) shows a gradient in oxygen, titanium and strontium as well as oxidization of the chromium interface layer while c) shows a constant oxygen concentration through the $SrTiO_3$ film. The gradient could originate from the $SrTiO_3$ reduction by chromium.

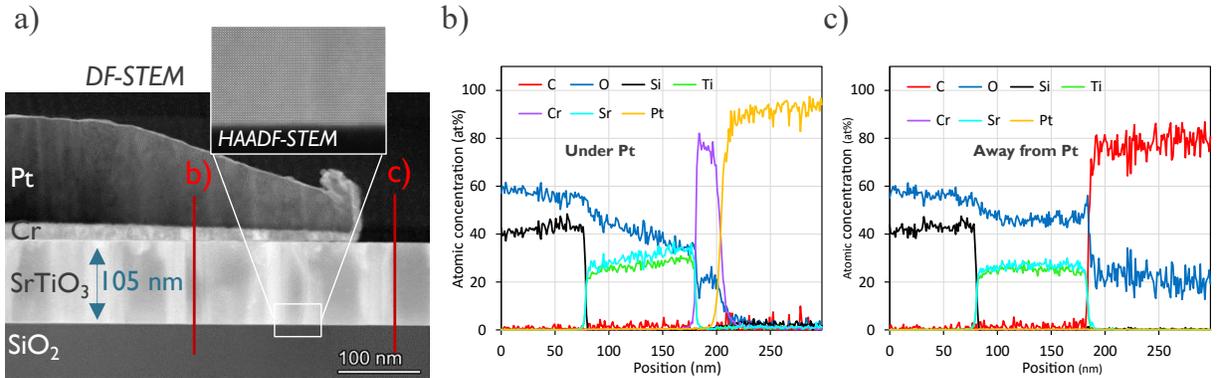

Supplementary Figure 12 a) Dark-field STEM image of the material stack at the electrodes of the EO device. Inset shows the HAADF-STEM of the SiO$_2$/SrTiO$_3$ interface. Red vertical lines show the position of the EDS of b) and c). b) EDS with 30 nm Cr & 140 nm Pt metal contacts. c) EDS without metal. The comparison of the two EDS shows that the oxygen and Sr, Ti profile is not flat as in c), which could be caused by the fully oxidized Cr.

## Supplementary 4.6. Room Temperature Measurements

Supplementary Figure 13 shows the optical output power versus three ac modulation periods at room temperature for an ac amplitude of 1 μm/V. The optical output power shows double the applied modulation frequency which is a clear indicator for a quadratic electro-optic Kerr effect; no Pockels effect is evident. The output signal can be fitted using equation (6) yielding a quadratic electro-optic Kerr coefficient of $R \approx 1.1 \cdot 10^{-17} \pm 1 \cdot 10^{-19}\ m^2/V^2$.

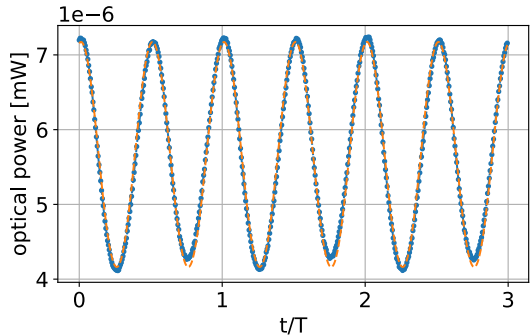

Supplementary Figure 13 The optical output power vs three periods of the ac modulation signal. The dashed line is the fit.

## Supplementary 5. Effective Pockels Coefficient Data

In the following we provide background information on the extraction procedure of the stated Pockels coefficients shown in Supplementary Figure 14 a) in the main text. We provide a motivation and the selection of references.

**OEO:** We are aware that certain OEO materials like pure, "bulk" JRD1 feature an electro-optic strength $n^3 \cdot r$ up to 2300 pm/V [18] or even 9000 pm/V for BAY1 (at 1310 nm) [19]. This is a larger value than the one indicated in the main text. However, these high Pockels coefficient in these materials go hand in hand with increased optical losses [19] which are undesired for quantum applications. Therefore, to do a more appropriate comparison, OEO materials featuring the lowest losses have been considered. These are also the materials that have been used by the community for cryogenic operation. The Pockels coefficient of OEO is driven by hyperpolarizability and not permittivity. As the main contribution to the permittivity is of electronic nature, only a small reduction in Pockels coefficient is expected when cooling down.

**LiNbO$_3$:** Bulk LiNbO$_3$ can be grown on wafer scale using different growth techniques such as the Czochralski method [20]. This combined with the development of the "smart cut" technology – ion slicing of bulk crystal and post wafer bonding onto an insulating substrate - enabled high-quality thin films.

**BaTiO$_3$:** The reference [21] in the main text reports Pockels values of $r_{eff} \approx 170$ pm/V corresponding to a third of its room temperature value $r_{eff} \approx 520$ pm/V. Highest Pockels values are achieved for full poling of the thin film. For this, the bias field needs to be above the coercive field of 8 V/µm (4 K), 0.5 V/µm (300 K). The reported permittivity values for different bias fields converge to a value of around 100-150 for 4 K. This corresponds to a reduction of roughly 80% compared to the room-temperature permittivity of 925 for the unbiased case (no bias data given for biases in between 0 and 4.8 V/µm). Following Miller's empirical rule, a cryogenic Pockels value of $r_{eff} \approx 20$ pm/V is expected. Recently, values of 13 pm/V at mK have been reported [22].

## Supplementary 6. Optical losses

Optical losses of the material are a crucial parameter for quantum applications. Supplementary Figure 14 shows the material loss of various Pockels materials. Cryogenic material loss of these materials has not yet been clearly reported which is why only room temperature values are compared here. The ferroelectric complex oxides have a bandgap of around 3-4 eV while the organic chromophores have a bandgap of around 1.5 eV in the visible. Thus, OEO commonly have a strong absorption peak around 850 nm wavelengths [23], which physically limits their material loss at 1550 nm in the telecommunication regime.

The material loss of 5.5 dB/cm in this work can be seen as an upper limit as it attributes the full propagation loss exclusively to the SrTiO$_3$, not considering other loss channels such as PMMA material loss or scattering loss [Supplementary 7]. Compared to LiNbO$_3$ with 0.2 dB/m, OEO with 3 dB/cm and BaTiO$_3$ with 0.2 dB/cm, our material is situated at the upper end in Supplementary Figure 14.

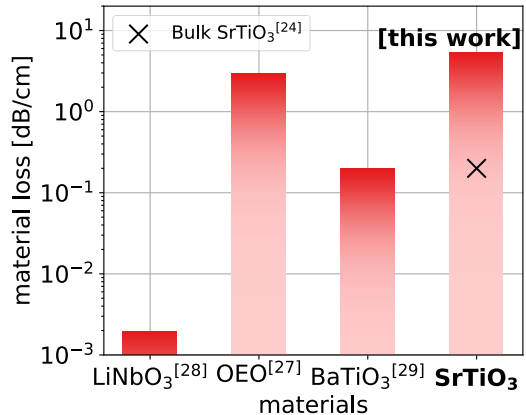

Supplementary Figure 14) Comparison of the optical material loss of different devices for room temperature at around 1550 nm wavelengths.

Further loss reduction efforts of our thin-film SrTiO$_3$ is needed to approach loss values observed in BaTiO$_3$ and bulk SrTiO$_3$ values of below 0.2 dB/cm [24].

In the following we provide background information on extraction of the stated number and motivation for the chosen references:

**OEO** For the OEO refence used in the main text [25] it is not possible to extract material loss of the used cross-linked HLD chromophore as a plasmonic structure was used as plasmonic propagation losses of dB/µm dominate. The datasheet of HLD states losses of ~ 35dB/cm at 1550 nm [26]. Note another, chromophore that reports a comparable Pockels coefficient and material losses of 3 dB/cm room temperature [27] has been used in Supplementary Figure 14. OEO materials with high Pockels but often also high optical losses are especially suitable for applications with short device lengths (i.e. plasmonic or slot waveguides) where their material loss contribution are negligible but currently not for quantum applications.

**LiNbO$_3$**: Post-fabrication treatments as annealing of thin films fabricated with the "smart cut" technology allows to partially restore the material loss of the bulk resulting in 2 dB/cm material loss [28].

**BaTiO$_3$**: Propagation loss values were extracted from reference [21]. However, this reference does not state the material losses explicitly and we provide here a first order estimation with the help of numerical simulations. The propagation losses of SiN waveguide loaded with BaTiO$_3$ and without BaTiO$_3$ are reported as 5.6 dB/cm, 5.4 dB/cm, respectively. Simulating the confinement in both cases, using the heights and geometrical design parameters, yielded a confinement of 0.24 in BaTiO$_3$, 0.33 in SiN and 0.3 in SiN without BaTiO$_3$. As the SiN confinement is almost the same in the case of with or without BaTiO$_3$, the material loss upper limit can be approximated by subtracting the two propagation losses, yielding:

$$\alpha_{prop,\text{BaTiO3}} \approx \frac{5.6 dB}{cm} = \Gamma_{BTO}\alpha_{\text{BaTiO3}} + \left(\overbrace{\Gamma_{SiN}}^{0.33} \alpha_{SiN} + \alpha_{scat}\right),$$

$$\alpha_{prop,\text{no BaTiO3}} \approx \frac{5.4 dB}{cm} \approx + \left(\overbrace{\Gamma_{SiN}}^{0.3} \alpha_{SiN} + \alpha_{scat}\right), \quad (26)$$

$$\rightarrow \alpha_{BtO} \leq \frac{\alpha_{prop,BTO} - \alpha_{prop,noBTO}}{\Gamma_{BTO}} \approx 1 dB/cm.$$

More recently, propagation loss values of ~ 0.1 dB/cm have been reported with a confinement of 63%, yielding a material loss of ~ 0.2 dB/cm [29].

## Supplementary 7. Optical loss measurement

Optical losses in the SrTiO$_3$ thin film were determined using a cut-back method at room temperature, complemented by substrate loss simulations. Light transmission through a series of PMMA waveguides, with lengths of 5 mm, 10 mm, 15 mm, and 20 mm, was measured using a fiber-to-fiber setup, as illustrated in Supplementary Figure 15 a). A tunable laser source, paired with a power meter, was used to investigate the optical properties across wavelengths ranging from 1500 nm to 1625 nm. Single-mode fibers coupled light to the chip via grating couplers. To isolate the losses attributable to SrTiO$_3$ from those of PMMA, waveguides were fabricated with different widths of 1.1 µm and 1.7 µm. For each waveguide length, three identical devices were measured. Propagation losses (see Supplementary Figure 15 c) for both waveguide widths were derived by fitting the transmission data (in dB) at each wavelength using a linear model across the four different lengths. The fitting was validated by correlation coefficient R values above 0.95. Losses from coupling to the Si substrate also contribute to the measured propagation losses. This was verified in Supplementary Figure 15 d), which illustrates the cross-sectional mode profile. The trend of the simulated substrate losses as a function of wavelength closely matches the one of the propagation losses. After subtracting the substrate losses, we obtain propagation losses of ~ 2 dB/cm for the 1.7 μm wide waveguides and lower losses of ~ 1.5 dB/cm for the 1.1 μm wide waveguide.

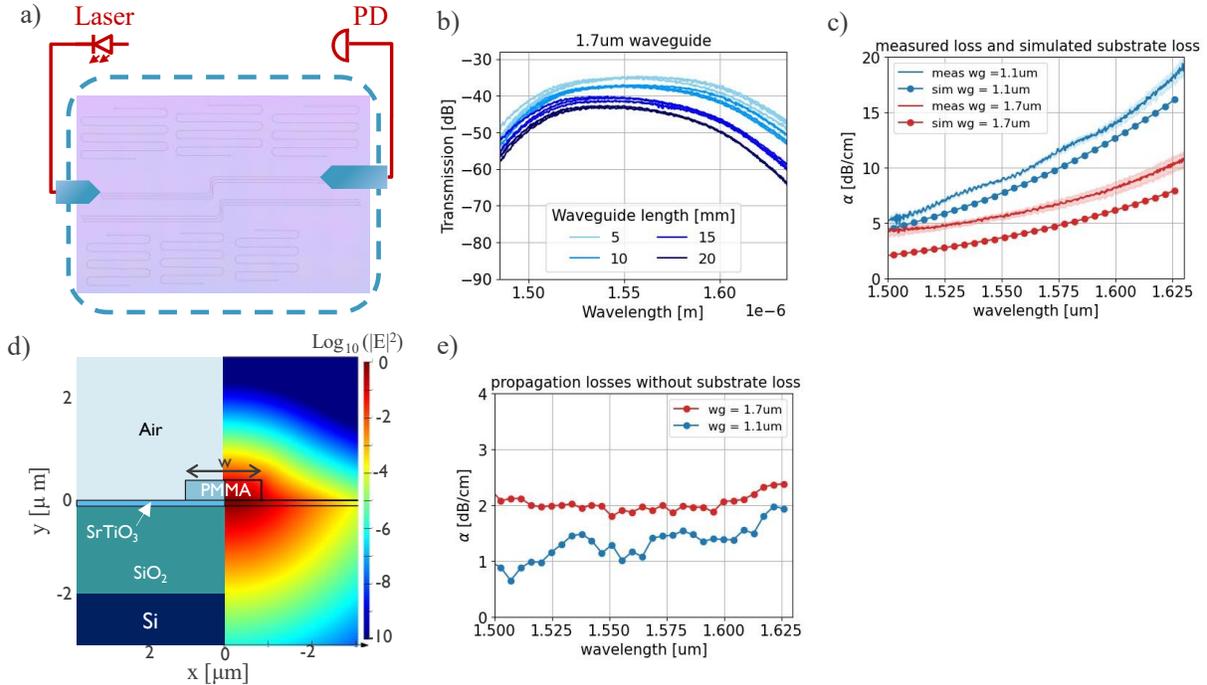

Supplementary Figure 15 a) Schematics of the cut-back measurement setup for optical losses. A tunable laser source, paired with a power meter, is used to measure the transmitted light via PMMA waveguides. b) Result of cut-back measurements for different waveguide lengths with 1.7 um width. c) Measured losses as a function of wavelength for both waveguide widths (shaded area around curves represents the standard deviation). The increasing losses with wavelength can be explained by the contribution of Si substrate losses (dots). d) Cross-sectional mode profile of the SrTiO3 sample, showing the Si substrate contribution. e) Propagation losses after subtracting substrate losses for both waveguide widths have an upper bound of ~ 2 dB/cm.

Light confinement within the $SrTiO_3$ is between 35% - 37% using the Poynting vector definition of confinement [9] for both widths. Thus, propagation losses should be equally for both types of waveguides. The difference between theory and experiment indicates that other loss mechanism could play a role. For instance, the light confinement within the PMMA increases for larger waveguide widths, which follows the scaling of the losses that increase with larger waveguide width. This would indicate significant loss contribution from the PMMA. However, further experiments with a larger variety of geometrical parameters and thicker $SiO_2$ substrates are needed to attribute the losses to individual sources. We identify 5.5 dB/cm as an upper material loss limit of the $SrTiO_3$ under the assumptions that $1^{st}$) all losses are caused by the $SrTiO_3$ and that $2^{nd}$) 37% of light is confined within the $SrTiO_3$ [9].

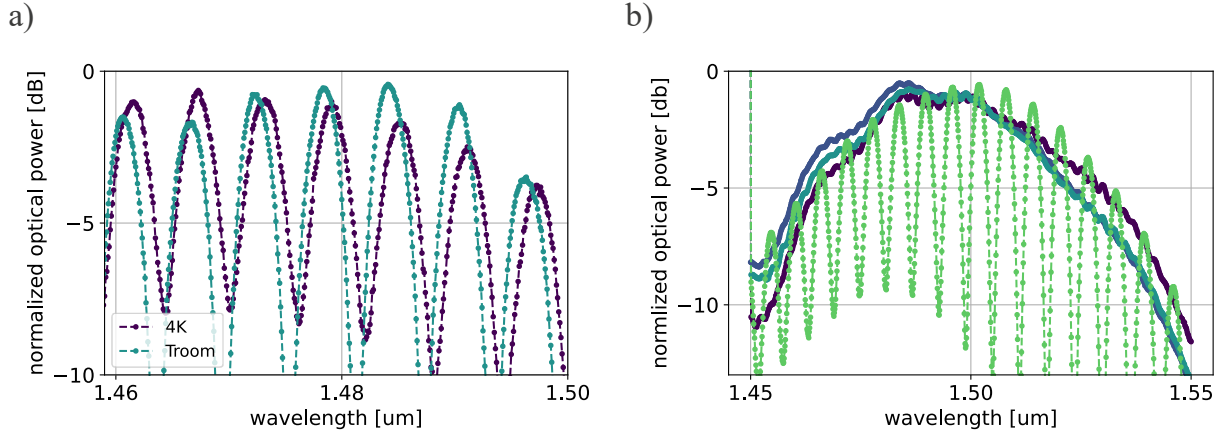

Supplementary Figure 16 a) Comparison of MZI spectrum at 4 K and room temperature. b): spectra of three waveguide with length of 5 mm (blue to purple) and spectrum of a MZI with arm length of 7.2 mm (green).

Supplementary Figure 16 a) shows a comparison of the 15.37 mm long MZI spectrum at room temperature and 4 K. While the peak of the optimal grating coupler is shifted to the blue for cryogenic temperatures, their power levels are comparable. Additionally, the group refractive index was extracted from the free-spectral range (FSR) of the seven constructive peaks using:

$$\Delta\lambda_{\text{FSR}} = \frac{\lambda_0^2}{n_g \Delta L}, \quad (27)$$

where $\Delta L = 200$ μm is the unbalance in the two arms. For room temperature this yields $1.86 \pm 0.1$ and for cryogenic $1.84 \pm 0.04$. A reduction in group refractive index would imply a slightly reduced refractive index and therefore reduced optical material loss according to Kramers-Kronig [15]. The reduction for cryogenic could as well be explained from the contraction of the PMMA and the reduction of confinement. The comparable power levels and the group refractive indices indicates that no significant loss mechanisms are active at cryogenic temperatures. Due to size constraint mentioned in S. Setup & Device, no cutback structures could be included in this design. However, another cryogenic measurement of three waveguides with 5 mm length and a MZI with 7.2 mm length shown in Supplementary Figure 16 b) was performed. The 2 mm difference in length introduces no strong losses and the power levels of the measurement stay within $< 0.5$ dB, as shown in Supplementary Figure 16 b). Therefore, if material loss increases at cryogenic temperatures, they stay below $(0.5 \text{ dB})/(0.37 \cdot 0.2 \text{ cm}) \approx 7 \text{ dB/cm}$. Further experiments are needed to precisely determine the introduced losses at cryogenic temperatures.

## Supplementary 8.   References

1. Boelen, A. *et al.* Stoichiometry and Thickness of Epitaxial SrTiO$_3$ on Silicon (001): an Investigation of Physical, Optical and Electrical Properties. Preprint at https://doi.org/10.48550/arXiv.2412.07395 (2024).
2. Haeni, J. H. *et al.* Room-temperature ferroelectricity in strained SrTiO$_3$. *Nature* **430**, 758–761 (2004).
3. Delhaye, G. *et al.* Structural properties of epitaxial SrTiO$_3$ thin films grown by molecular beam epitaxy on Si(001). *J. Appl. Phys.* **100**, 124109 (2006).

4. Carneiro, J. O., Teixeira, V. & Azevedo, S. Residual Stresses in Thin Films Evaluated by Different Experimental Techniques. in *Encyclopedia of Thermal Stresses* (ed. Hetnarski, R. B.) 4222–4231 (Springer Netherlands, Dordrecht, 2014). doi:10.1007/978-94-007-2739-7_64.
5. Dwij, V., De, B. K., Tyagi, S., Sharma, G. & Sathe, V. Fano resonance and relaxor behavior in Pr doped SrTiO3: A Raman spectroscopic investigation. *Phys. B Condens. Matter* **620**, 413265 (2021).
6. Linnik, E. D. *et al.* Raman Response of Quantum Critical Ferroelectric Pb-Doped SrTiO3. *Crystals* **11**, 1469 (2021).
7. Tenne, D. A. *et al.* Ferroelectricity in nonstoichiometric SrTiO3 films studied by ultraviolet Raman spectroscopy. *Appl. Phys. Lett.* **97**, 142901 (2010).
8. Haffner, C. *et al.* All-plasmonic Mach–Zehnder modulator enabling optical high-speed communication at the microscale. *Nat. Photonics* **9**, 525–528 (2015).
9. Aellen, M. & Norris, D. J. Understanding Optical Gain: Which Confinement Factor is Correct? *ACS Photonics* **9**, 3498–3505 (2022).
10. Ma, Z. *et al.* Modeling of hysteresis loop and its applications in ferroelectric materials. *Ceram. Int.* **44**, 4338–4343 (2018).
11. Wördenweber, R., Schubert, J., Ehlig, T. & Hollmann, E. Relaxor ferro- and paraelectricity in anisotropically strained SrTiO3 films. *J. Appl. Phys.* **113**, 164103 (2013).
12. Li, Y. L. *et al.* Phase transitions and domain structures in strained pseudocubic (100) Sr Ti O 3 thin films. *Phys. Rev. B* **73**, 184112 (2006).
13. Xu, R. *et al.* Strain-induced room-temperature ferroelectricity in SrTiO3 membranes. *Nat. Commun.* **11**, 3141 (2020).
14. Chelladurai, D. *et al.* Barium Titanate and Lithium Niobate Permittivity and Pockels Coefficients from MHz to Sub-THz Frequencies. Preprint at https://doi.org/10.48550/arXiv.2407.03443 (2024).
15. Boyd, R. W. *Nonlinear Optics*. (Academic Press, Burlington, MA, 2008).
16. Weilenmann, C. *et al.* Single neuromorphic memristor closely emulates multiple synaptic mechanisms for energy efficient neural networks. *Nat. Commun.* **15**, 6898 (2024).
17. De Souza, R. A. Oxygen Diffusion in SrTiO3 and Related Perovskite Oxides. *Adv. Funct. Mater.* **25**, 6326–6342 (2015).
18. Kieninger, C. *et al.* Ultra-high electro-optic activity demonstrated in a silicon-organic hybrid modulator. *Optica* **5**, 739 (2018).
19. Xu, H. *et al.* Electro-Optic Activity in Excess of 1000 pm V$^{-1}$ Achieved via Theory-Guided Organic Chromophore Design. *Adv. Mater.* **33**, 2104174 (2021).
20. Boes, A. *et al.* Lithium niobate photonics: Unlocking the electromagnetic spectrum. *Science* **379**, eabj4396 (2023).
21. Eltes, F. *et al.* An integrated optical modulator operating at cryogenic temperatures. *Nat. Mater.* **19**, 1164–1168 (2020).
22. Möhl, C. *et al.* Bidirectional microwave-optical conversion using an integrated barium-titanate transducer. Preprint at https://doi.org/10.48550/arXiv.2501.09728 (2025).
23. Feng, S., Wu, S., Zhang, W., Liu, F. & Wang, J. Organic Electro-Optic Materials with High Electro-Optic Coefficients and Strong Stability. *Molecules* **29**, 3188 (2024).
24. Anderson, C. P. et al. Quantum critical electro-optic and piezo-electric nonlinearities. Preprint at https://doi.org/10.48550/arXiv.2502.15164 (2025).
25. Bisang, D. *et al.* Plasmonic Modulators in Cryogenic Environment Featuring Bandwidths in Excess of 100 GHz and Reduced Plasmonic Losses. *ACS Photonics* **11**, 2691–2699 (2024).

26. HLD - Organic Non-linear Optical (NLO) Material with High Electro-optic Effects. (2021).
27. Lu, G.-W. *et al.* High-temperature-resistant silicon-polymer hybrid modulator operating at up to 200 Gbit s−1 for energy-efficient datacentres and harsh-environment applications. *Nat. Commun.* **11**, 4224 (2020).
28. Shams-Ansari, A. *et al.* Reduced material loss in thin-film lithium niobate waveguides. *APL Photonics* **7**, 081301 (2022).
29. Riedhauser, A. *et al.* Absorption loss and Kerr nonlinearity in barium titanate waveguides. *APL Photonics* **10**, 016121 (2025).